\documentclass[aps,amssymb,amsmath,
twocolumn,showpacs,superscriptaddress,
groupedaddress,ifmbe]{revtex4-1}

\usepackage{graphicx}
\usepackage{natbib}
\usepackage{dcolumn}
\usepackage{bm}
\usepackage[table,xcdraw]{xcolor}
\usepackage{nth}
\usepackage[ruled,vlined]{algorithm2e}
\usepackage{framed}
\usepackage{listings}
\usepackage[titletoc,title]{appendix}
\usepackage{tikz}
\usepackage{lipsum}
\usepackage{hyperref}
\usepackage{multirow}
\usepackage{nicefrac}
\usepackage[table]{xcolor}
\usepackage{threeparttable}
\usepackage{comment}
\usepackage{amsmath}
\usepackage[autostyle]{csquotes}
\usepackage{float}
\usepackage{placeins}

\newcommand*{\dprime}{^{\prime\prime}\mkern-1.2mu}

\hypersetup{
  colorlinks,
  citecolor=red,
  linkcolor=blue,
  urlcolor=blue}

\begin{document}
\title{A treatment of $d$-dimensional Coulomb$-$Sturmians with noninteger quantum numbers and their connection to the radial momentum operator}
\author{Ali Ba{\u g}c{\i}}
\email{abagci@pau.edu.tr}
\affiliation{Computational and Gravitational Physics Laboratory, Department of Physics, Faculty of Science, Pamukkale University, Denizli, Turkey}
\author{Z{\" u}hra G{\" u}ne{\c s}}
\affiliation{Computational and Gravitational Physics Laboratory, Department of Physics, Faculty of Science, Pamukkale University, Denizli, Turkey}
\begin{abstract}
The Ba\u{g}c\i$-$Hoggan exponential$-$type orbitals, the generalization of Coulomb$-$Sturmians to noninteger quantum numbers, were recently identified as $d-$dimensional Coulomb$-$Sturmians. In the present study, this identification is treated in detail. The radial differential equation for these orbitals is derived directly from their closed form. It represents the general Coulomb$-$Sturmian differential equation for quantum numbers with real values. The solution is verified through a power series representation and a first$-$order matrix form based on the Green's function of the radial Laplacian. New transformation relationships between these orbitals and Slater$-$type orbitals with noninteger principal quantum numbers are established. The radial differential equation is expressed in terms of the radial momentum operator. The expectation values of the position and radial momentum operators are obtained in closed form. The radial and Heisenberg uncertainty products, the Shannon entropy and the Fisher information are derived in general form.
\begin{description}
\item[Keywords]
Ba\u{g}c\i$-$Hoggan ETOs, Coulomb$-$Sturmian functions, radial momentum operator
\end{description}
\end{abstract}
\maketitle
\section{Introduction} \label{sec:intro}
Hydrogen$-$like wave functions are exact solutions of the time$-$independent Schr{\"o}dinger equation for an electron moving in the Coulomb potential of a fixed, point$-$like nucleus~\cite{1Condon1929,2Galler2021}, making them suitable candidates for the construction of atomic basis sets. Their discrete set, however, is not complete without the continuum states. This incompleteness was first overcome by Hylleraas~\cite{3Hylleraas1928}, who used a single orbital exponent $\zeta$ for all functions of a set. The resulting functions were later termed Sturmians by Rotenberg~\cite{4Rotenberg1962}. They are solutions of a Sturm$-$Liouville problem~\cite{4Rotenberg1962,5Rotenberg1970}. In contrast to the hydrogen$-$like case, the energy here is fixed and common to all functions of the set. The eigenvalue is instead the factor multiplying the Coulomb potential. These eigenvalues form an infinite, discrete and strictly increasing sequence. They are orthonormal with respect to a weight containing the Coulomb potential and form a complete discrete set, both in the Hilbert space~\cite{6Klahn1977,7Klahn1977} and in the Sobolev space~\cite{6Klahn1977,7Klahn1977,8Sobolev1963,9Weniger1985,10Weniger2008,11Weniger2012}.\\
Coulomb$-$Sturmians satisfy the nuclear cusp condition, exhibit the correct exponential decay at large distances~\cite{12Kato1957,13Agmon1982} and possess radial nodes. These properties make them suitable not only for electronic structure of atoms and molecules~\cite{14Avery2003,15Herbst2019} but also for the calculation of nuclear magnetic resonance shielding tensors~\cite{16Hoggan2009} and for the description of Rydberg$-$type and autoionizing states~\cite{17Ambrosio2014}. Their exponential decay also makes them an efficient basis beyond atomic physics: in nuclear many$-$body calculations~\cite{18Caprio2012}, in particular for halo nuclei such as $^{6}$He and $^{8}$He~\cite{19Caprio2014}, and in the description of Rydberg excitons confined in finite crystals~\cite{20Belov2024}. The Coulomb problem from which they originate arises in other fields as well. The Wheeler$-$DeWitt equation of a flat universe takes the form of the radial hydrogen$-$like equation, with an angular parameter that need not be an integer~\cite{21Mukherjee2026}. An ultralight boson, analogously to an electron bound in the Coulomb potential of an atomic nucleus, is bound in the gravitational potential of a black hole. This forms a system known as a gravitational atom. In the limit of weak gravitational coupling, the energy levels of such systems exhibit a hydrogen$-$like spectrum~\cite{22Arvanitaki2011, 23Baumann2019, 24DellaRocca2026}. Their $d-$dimensional representation further arises in quantum field theory~\cite{25Itzykson1980,26Dong2002} and quantum cosmology~\cite{21Mukherjee2026,27AmelinoCamelia2005}.

Coulomb$-$Sturmians are directly connected to Slater$-$type orbitals (STOs)~\cite{28Slater1930}. STOs were introduced empirically and consist of a single power of the radial coordinate and an exponential factor. They contain no polynomial factor and therefore have no radial nodes. Coulomb$-$Sturmians differ from STOs by a Laguerre polynomial, so that each Coulomb$-$Sturmian is a finite linear combination of STOs with a common exponent~\cite{9Weniger1985, 29Steinborn1982,30Calderini2012}. The Laguerre polynomials make the use of Coulomb$-$Sturmians in atomic and molecular structure calculations and in the evaluation of physical properties more complicated~\cite{29Steinborn1982,31Guseinov1970,32Mamedov2014}, particularly for noninteger quantum numbers~\cite{10Weniger2008,11Weniger2012,33Weniger2021}.\\
Quantum numbers with noninteger values have been adopted in atomic physics from three different standpoints. The first is the variational standpoint. The principal quantum number with noninteger values was first considered in calculations of many$-$electron atoms~\cite{34Zener1930,35Slater1930}. Later, it was treated as a parameter in the STOs to be determined variationally~\cite{36Parr1957}. Such calculations do not require the basis functions to be solutions of a differential equation, but only to be square$-$integrable and to produce finite expectation values of the Hamiltonian. Slater$-$type orbitals with noninteger principal quantum numbers ($n^{\ast}-$STOs) have subsequently been applied in numerous Hartree$-$Fock calculations for atoms. For a comprehensive historical account of their development and applications, the reader is referred to~\cite{37Bagci2020} and the references therein.

The second is spectroscopic. The energy levels of a single excited electron outside a closed atomic core are described by an effective noninteger principal quantum number~\cite{38Seaton1983}
\begin{align}\label{eq:s11}
E_{n}&=-\frac{Z^{\prime 2}}
{2\left(n-\epsilon\right)^{2}},
&
n^{*}_{\epsilon}=n-\epsilon,
\end{align}
where $\epsilon$ is the quantum defect and $Z^{\prime}$ is the net charge of the core. The quantum defect arises from the penetration of the electron into the core, where the nucleus is no longer screened completely, and from the polarization of the core by the electron. Outside the core, the electron moves in the Coulomb potential of the net charge $Z^{\prime}$, as in the hydrogen$-$like ion, so that its radial function is a combination of the linearly independent Coulomb functions $f$ and $g$. The function $f$ is regular at the origin, and the function $g$ is irregular there~\cite{38Seaton1983}. For $r>r_{0}$, where $r_{0}$ is the radius of the core,
\begin{align}\label{eq:s12}
F\left(E_{n},r\right)
=\cos\left(\pi\epsilon\right)f\left(E_{n},r\right)
+\sin\left(\pi\epsilon\right)g\left(E_{n},r\right).
\end{align}
In the hydrogen atom only the regular function $f$ is allowed and $\epsilon=0$. In many$-$electron atoms, the Coulomb form of the potential holds only beyond $r_{0}$, so that the regularity condition at the origin no longer applies to the outer solution, and the irregular function $g$ also contributes. The requirement that the function of Eq.~\eqref{eq:s12} decays at large distances leads to the noninteger effective quantum number $n^{\ast}_{\epsilon}$. A few parameters, therefore, describe an entire series of bound states together with the continuum states~\cite{38Seaton1983,39Fu2016}. Quantum defect theory is thus applied where many$-$electron calculations become impractical. Highly excited states are obtained by extrapolation from directly calculated ones~\cite{40Drake2026}, ionization energies are determined with the highest precision~\cite{41Mokhberi2019,42Shen2025}, and the same description applies to molecules~\cite{43Doran2026} and to excitons in semiconductors~\cite{44Ertl2026}.
When the effect of the core is represented by an additional term in the centrifugal part of the radial differential equation, its solution is obtained in hydrogen$-$like form. Both the principal and the angular momentum quantum numbers then take noninteger values, $n^{\ast}=n-\delta(l)$ and $l^{\ast}=l-\delta(l)+I(l)$, where $I(l)$ is an integer, so that $n^{\ast}-l^{\ast}-1=n-l-1-I(l)$ remains an integer and the solutions retain the hydrogen$-$like form with Laguerre polynomials~\cite{45Kostelecky1985}. In the weakest$-$bound$-$electron potential model, an analogous form, $n^{\ast}=n+\epsilon$ and $l^{\ast}=l+\epsilon$, is obtained, where $\epsilon$ is here a parameter of the effective potential~\cite{46Fatima2024}. Such solutions of the radial differential equation for noninteger quantum numbers constitute the third standpoint.

Infeld and Hull~\cite{47Infeld1951} showed that such quantum numbers emerge in the solution of the generalized Kepler problem, and they also appear in the Dirac$-$Coulomb problem through $\gamma=\sqrt{\kappa^{2}-Z^{2}/c^{2}}$. Here, $\kappa$ is the eigenvalue of the Dirac spin$-$orbit operator~\cite{48Greiner2000}. Infeld and Hull used the factorization method for the solution of the differential equations. The solutions were represented in terms of Slater$-$type orbitals with noninteger principal quantum numbers ($n^{\ast}_{\epsilon}-$STOs) via recurrence relationships. It should be noted that the $n^{\ast}-$STOs differ from the $n^{\ast}_{\epsilon}-$STOs. In the former, the principal quantum number $n^{\ast}$ is a free variational parameter, restricted only by $n^{\ast}>0$. In the latter, $n^{\ast}_{\epsilon}=n+\epsilon$ is an integer $n$ shifted by a noninteger value $\epsilon$, with $n^{\ast}_{\epsilon}\ge1$. In quantum defect theory, the shift is toward lower values, $n^{\ast}_{\epsilon}=n-\epsilon$, and $n^{\ast}_{\epsilon}$ is determined by the measured energy, whereas the integer $n$ serves only to label the levels of a series. Since the binding energy of an electron outside a closed core does not exceed that of the ground state of the hydrogen$-$like ion with the same net charge, $n^{\ast}_{\epsilon}\ge1$ also holds in this case.\\
The solutions provided by Infeld and Hull constitute a Coulomb$-$Sturmian$-$like set with noninteger quantum numbers, from which the $n^{\ast}_{\epsilon}-$STOs are obtained by retaining only the highest power of $r$ in the Laguerre polynomials. For the $n^{\ast}-$STOs employed in variational calculations, with $n^{\ast}>0$, however, a complete orthonormal set from which they could be derived was absent. This problem was solved by one of the present authors. By means of the Riemann$-$Liouville fractional calculus~\cite{49Kilbas2006}, complete and orthonormal sets of exponential$-$type orbitals with noninteger quantum numbers were obtained for $n^{\ast}>0$~\cite{50Bagci2023}. These functions, hereafter referred to as BH$-$ETOs~\cite{51Bagci2025,52Bagci2025}, reduce to the solutions of Infeld and Hull for $n^{\ast}\ge 1$, and to the $n^{\ast}-$STOs under the same simplification of their Laguerre polynomials.\\
The identification of the BH$-$ETOs as $d-$dimensional Coulomb$-$Sturmians with noninteger quantum numbers was subsequently reported in a technical note~\cite{53Bagci2026}. There, the variable$-$dependent and constant terms of the differential equation of the $\Psi^{\alpha}-$ETOs, complete orthonormal sets of exponential$-$type orbitals with integer quantum numbers defined in weighted Hilbert spaces~\cite{54Guseinov2002}, were properly separated. The resulting equation has the Sturm$-$Liouville form of the $d-$dimensional Coulomb$-$Sturmian problem, in which the parameter $\alpha=4-d$ is fixed by the dimension. The differential equation of the BH$-$ETOs, written by analogy, has the same form and reduces to that of the Coulomb$-$Sturmians for $\nu=1$ and integer values of $n^{\ast}$ and $l^{\ast}$, where $0< \nu \leq 1$ is the parameter introduced in Section~\ref{sec:BHETOsDDim}. The BH$-$ETOs thus constitute the generalization of the Coulomb$-$Sturmians to noninteger quantum numbers.

In the present study, the BH$-$ETOs are treated as $d-$dimensional Coulomb$-$Sturmians with noninteger quantum numbers. A detailed derivation of their differential equation and its solution, which lay beyond the scope of that technical note, is presented here. In addition, the radial differential equation is expressed in terms of the radial momentum operator, and the expectation values and uncertainty measures of the BH$-$ETOs are obtained.\\
The paper is organized as follows. In Section~\ref{sec:GenRadEq}, the radial differential equation is derived explicitly from the closed form of the BH$-$ETOs. This is done without mixing the centrifugal and Coulomb contributions. The $\Psi^{\alpha}-$ETOs and BH$-$ETOs  are discussed as $d-$dimensional Coulomb$-$Sturmians in Sections~\ref{sec:PAETOsDDim},~\ref{sec:BHETOsDDim}, respectively. In Section~\ref{sec:BHETOsDifSol}, the solution is verified analytically through a power series representation (Section~\ref{sec:BHETOsSeries}) and, alternatively, through a first$-$order matrix differential equation (Section ~\ref{sec:BHETOsFirstOrd}) obtained from the Green's function of the radial Laplacian. In Section~\ref{sec:OptFRadDif}, the radial differential equation is written in terms of the radial momentum operator, and the corresponding Hamiltonian is presented. New transformation relationships between the BH$-$ETOs and $d-$dimensional Slater$-$type orbitals with noninteger quantum numbers are derived, and through them the expectation values of the position and radial momentum operators are obtained in closed form. In Section~\ref{sec:UncertainityBHETOs}, the radial and Heisenberg uncertainty products, the Shannon entropy and the Fisher information of the BH$-$ETOs are given in general form. Section~\ref{sec:Conclusions} contains the conclusion.

Throughout this work, the derivation is formulated in terms of the energy parameter $E$, introduced in the Coulomb$-$Sturmian form of the radial differential equation for mathematical convenience. The parameter $E$ is distinct from the physical energy, defined as the eigenvalue of the Hamiltonian. The two are related by a constant factor. Atomic units are used unless otherwise stated.
\section{Generalized radial equation}\label{sec:GenRadEq}
A proper separation of the variable$-$dependent and constant terms is essential in deriving the radial differential equation of the Coulomb$-$Sturmians. Only then does the equation retain the Sturm$-$Liouville form that defines these functions. In this form, the radial operator is self$-$adjoint with respect to the volume element $r^{d-1}dr$ of a $d-$dimensional space. The dimension $d$ can therefore be directly read from the differential equation, through the coefficient $(d-1)/r$ of its first$-$derivative term. If the centrifugal and Coulomb terms are mixed, this is no longer the case. The equation then takes a three$-$dimensional form with an additional term, and its radial operator is not self$-$adjoint with respect to the three$-$dimensional volume element $r^{2}dr$. Orthogonality must then be imposed through a weight in the inner product, and the dimension determined by the differential equation is obscured.

The differential equation of the BH$-$ETOs is derived explicitly from their closed form. The $d-$dimensional BH$-$ETOs, with an appropriate substitution of parameters, can be expressed as,
\begin{align}\label{eq:s23}
\mathcal{R}^{\alpha}_{n^{\ast}l^{\ast}}
\left(\zeta, r\right)
=\mathcal{N}^{\alpha}_{n^{\ast}l^{\ast}}\left(\zeta\right)
x^{\beta^{\ast}}
e^{-x/2}
\mathcal{L}^{p^{\ast}}_{q^{\ast}-p^{\ast}}\left(x\right)
\end{align}
here, $\alpha=4-d$, $\beta^{*}=(p^{*}+\alpha)/2-1$ and $x=2\zeta r$. The normalization constant with respect to the volume element $r^{3-\alpha}dr$ of the $d-$dimensional space is,
\begin{multline}\label{eq:s24}
\mathcal{N}^{\alpha}_{n^{\ast}l^{\ast}}\left(\zeta\right)
=\left(
\frac{\left(2\zeta\right)^{4-\alpha}}
{2q^{\ast}-p^{\ast}+1}
\right)^{1/2}
\\
\times
\left[
\frac{\Gamma\left(q^{\ast}-p^{\ast}+1\right)}
{\Gamma\left(q^{\ast}+1\right)}
\right]^{1/2},
\end{multline}
where, $\Gamma(x)$ is the gamma function. Although the differential equations of the BH$-$ETOs are expressed in terms of the parameter $\alpha$, as in the factor $\left(3-\alpha\right)$, the space in which they are defined, in accordance with the convention adopted in the literature, is referred to throughout the present paper as $d-$dimensional.\\
Writing Eq.~\eqref{eq:s23} as a product of the factors $u=x^{\beta^{\ast}}$, $v=e^{-x/2}$ and $\omega=\mathcal{L}^{p^{\ast}}_{q^{\ast}-p^{\ast}}$, the first derivative with respect to $x$ is readily obtained as,
\begin{multline}\label{eq:s25}
\mathcal{R}^{\prime\alpha}_{n^{\ast}l^{\ast}}\left(\zeta, r\right)
=\mathcal{N}^{\alpha}_{n^{\ast}l^{\ast}}\left(\zeta\right)
x^{\beta^{\ast}}
e^{-x/2}
\Big(
\frac{\beta^{\ast}}{x}
\mathcal{L}^{p^{\ast}}_{q^{\ast}-p^{\ast}}\left(x\right)
\\
-\frac{1}{2}
\mathcal{L}^{p^{\ast}}_{q^{\ast}-p^{\ast}}\left(x\right)
+\mathcal{L}^{\prime p^{\ast}}_{q^{\ast}-p^{\ast}}\left(x\right)
\Big),
\end{multline}
where the prime denotes the derivative with respect to $x$, $d/dr=2\zeta\,d/dx$, and,
\begin{multline}\label{eq:s26}
\mathcal{R}^{\prime\alpha}_{n^{\ast}l^{\ast}}\left(\zeta, r\right)
=\frac{\beta^{\ast}}{x}
\mathcal{R}^{\alpha}_{n^{\ast}l^{\ast}}\left(\zeta, r\right)
-\frac{1}{2}
\mathcal{R}^{\alpha}_{n^{\ast}l^{\ast}}\left(\zeta, r\right)
\\
+\mathcal{N}^{\alpha}_{n^{\ast}l^{\ast}}\left(\zeta\right)
x^{\beta^{\ast}}
e^{-x/2}
\mathcal{L}^{\prime p^{\ast}}_{q^{\ast}-p^{\ast}}\left(x\right).
\end{multline}
Differentiating Eq.~\eqref{eq:s25} once more and using the differential equation of the generalized Laguerre polynomials~\cite{55Magnus1966},
\begin{multline}\label{eq:s27}
\mathcal{L}^{\dprime p^{\ast}}_{q^{\ast}-p^{\ast}}\left(x\right)
=-\frac{\left(p^{\ast}+1-x\right)}{x}
\mathcal{L}^{\prime p^{\ast}}_{q^{\ast}-p^{\ast}}\left(x\right)
\\
-\frac{\left(q^{\ast}-p^{\ast}\right)}{x}
\mathcal{L}^{p^{\ast}}_{q^{\ast}-p^{\ast}}\left(x\right),
\end{multline}
the second derivative of the BH$-$ETOs takes the form,
\begin{multline}\label{eq:s28}
\mathcal{R}^{\dprime\alpha}_{n^{\ast}l^{\ast}}\left(\zeta, r\right)
=\mathcal{N}^{\alpha}_{n^{\ast}l^{\ast}}\left(\zeta\right)
x^{\beta^{\ast}}e^{-x/2}
\\
\times
\Bigg[
\left(
\frac{\beta^{\ast}\left(\beta^{\ast}-1\right)}{x^{2}}
-\frac{\beta^{\ast}}{x}+\frac{1}{4}
-\frac{\left(q^{\ast}-p^{\ast}\right)}{x}\right)
\mathcal{L}^{p^{\ast}}_{q^{\ast}-p^{\ast}}\left(x\right)
\\
+\left(
\frac{2\beta^{\ast}}{x}
-1
-\frac{\left(p^{\ast}+1-x\right)}{x}
\right)
\mathcal{L}^{\prime p^{\ast}}_{q^{\ast}-p^{\ast}}\left(x\right)\Bigg].
\end{multline}
Extracting the term involving the Laguerre polynomials from Eq.~\eqref{eq:s26} and substituting
it into Eq.~\eqref{eq:s28}, the second derivative is obtained in the form,
\begin{multline}\label{eq:s29}
\mathcal{R}^{\dprime\alpha}_{n^{\ast}l^{\ast}}\left(\zeta, r\right)
=B^{\beta^{\ast}}_{q^{\ast}-p^{\ast}}
\left(x\right)
\mathcal{R}^{\prime\alpha}_{n^{\ast}l^{\ast}}\left(\zeta, r\right)
\\
+\left[
A^{\beta^{\ast}}_{q^{\ast}-p^{\ast}}\left(x\right)
+B^{\beta^{\ast}}_{q^{\ast}-p^{\ast}}
\left(x\right)
\left(
\frac{1}{2}-\frac{\beta^{\ast}}{x}
\right)
\right]
\\
\times
\mathcal{R}^{\alpha}_{n^{\ast}l^{\ast}}\left(\zeta, r\right),
\end{multline}
where $A^{\beta^{\ast}}_{q^{*}-p^{\ast}}(x)$ and $B^{\beta^{\ast}}_{q^{\ast}-p^{\ast}}(x)$ stand for the parenthesized expressions in the first and second terms of Eq.~\eqref{eq:s28}, respectively. A straightforward rearrangement of the bracketed terms in Eq.~\eqref{eq:s29}, together with the definition of the parameter $\beta^{\ast}$ given below Eq.~\eqref{eq:s23}, yields,
\begin{multline}\label{eq:s210}
x\mathcal{R}^{\dprime\alpha}_{n^{\ast}l^{\ast}}\left(\zeta, r\right)
+\left(3-\alpha\right)
\mathcal{R}^{\prime\alpha}_{n^{\ast}l^{\ast}}\left(\zeta, r\right)
\\
-\frac{D^{p^{\ast}}_{q^{\ast}-p^{\ast}}\left(x\right)}{4x}
\mathcal{R}^{\alpha}_{n^{\ast}l^{\ast}}\left(\zeta, r\right)=0.
\end{multline}
The constant term of the function $D^{p^{\ast}}_{q^{\ast}-p^{\ast}}$ contains the eigenvalue of the angular operator and the function is of particular significance for the interpretation of differential equations of the form of Eq.~\eqref{eq:s210}. Note that such equations admit Laguerre$-$type solutions as given in Eq.~\eqref{eq:s23} when $D^{p^{\ast}}_{q^{\ast}-p^{\ast}}$ is a polynomial of at most second degree~\cite{56Nikiforov1988}. The function $D^{p^{\ast}}_{q^{\ast}-p^{\ast}}$ used here reads:
\begin{multline}\label{eq:s211}
D^{p^{\ast}}_{q^{\ast}-p^{\ast}}\left(x\right)
\\
=p^{\ast 2}-\alpha^{2}+4\alpha-4+2x\left(p^{\ast}-2q^{\ast}-1\right)+x^{2}.
\end{multline}
In the subsequent sections only the values of $p^{\ast},q^{\ast}$ are given, as these uniquely determine the explicit form of the solution.
\subsection{$\Psi^{\alpha}-$ETOs and $d-$Dimensional Coulomb$-$Sturmian Functions}\label{sec:PAETOsDDim}
The $\Psi^{\alpha}-$ETOs, with $n\in\mathbb N_{>0}$, $l\in\{0,1,\ldots,n-1\}$, $-l\le m\le l$, were defined by Guseinov~\cite{54Guseinov2002}. These functions were treated~\cite{54Guseinov2002,57Guseinov2007} as elements of a weighted Hilbert space~\cite{6Klahn1977,10Weniger2008,11Weniger2012}. They are characterized by the values of the parameters $p^{\ast},q^{\ast}$ given as $p^{\ast}=p=2l+2-\alpha$, $q^{\ast}-p^{\ast}=q-p=n-l-1$, $\{p,q\}\in\mathbb N_{\ge0}$. The differential equation satisfied by these functions was given as~\cite{58Guseinov2012},
\begin{multline}\label{eq:s212}
x\frac{d^{2}}{dx^{2}}
\mathcal{R}^{\alpha}_{nl}\left(x\right)
+\left(3-\alpha\right)\frac{d}{dx}
\mathcal{R}^{\alpha}_{nl}\left(x\right)
\\
+\left[
n+\left(1-\alpha\right)
\left(\frac{1}{2}-\frac{l}{x}\right)
\right.
\\
\left.
-\frac{l\left(l+1\right)}{x}
-\frac{x}{4}
\right]
\mathcal{R}^{\alpha}_{nl}\left(x\right)=0.
\end{multline}
Eq.~\eqref{eq:s212} governing the $\Psi^{\alpha}-$ETOs is obtained by the full reduction of the function $D^{p}_{q-p}$. It is expressed in this particular form in order to retain the centrifugal term resulting from the eigenvalue of the squared orbital angular momentum operator acting on the three$-$dimensional spherical harmonics. This kind of reduction, although mathematically legitimate, is physically improper. It mixes centrifugal and Coulomb contributions and thereby obscures the Coulomb$-$Sturmian structure~\cite{53Bagci2026} (see Eqs.~\eqref{eq:s213}$-$\eqref{eq:s216}).

The origin of this mixing of centrifugal and Coulomb contributions in Eq.~\eqref{eq:s212} becomes transparent when it is rewritten so that the three$-$dimensional Laplacian appears as one complete block. Splitting the coefficient of the first derivative as $3-\alpha=2+(1-\alpha)$, expanding the factor $(1-\alpha)\left(\frac12-\frac lx\right)$ and returning to the variable $r=x/\left(2\zeta\right)$, Eq.~\eqref{eq:s212} written in the three$-$dimensional form with $\mathcal{R}^{\alpha}_{nl}(\zeta,r)$ replaced by $R^{\alpha}_{nl}(\zeta,r)$ and
\begin{align*}
\Psi^{\alpha}_{nlm}(\zeta,\vec r)=
R^{\alpha}_{nl}(\zeta,r)\,Y_{lm}(\theta,\varphi)
\end{align*}
takes the form,
\begin{multline}\label{eq:s213}
\nabla^{2}_{2}
\Psi^{\alpha}_{nlm}\left(\zeta, \vec{r}\right)
+\frac{1-\alpha}{r}
\left(
\frac{\partial}{\partial r}-\frac{l}{r}
\right)
\Psi^{\alpha}_{nlm}\left(\zeta, \vec{r}\right)
\\
+\left[
\frac{\zeta\left(2n+1-\alpha\right)}{r}-\zeta^{2}
\right]
\Psi^{\alpha}_{nlm}\left(\zeta, \vec{r}\right)=0,
\end{multline}
where $\nabla^{2}_{2}$ denotes the three$-$dimensional Laplacian. The subscript gives the dimension of the unit sphere. The second term, as an additional operator, is the price of retaining the three$-$dimensional centrifugal term. It contains the quantum number $l$ explicitly. The $\Psi^{\alpha}-$ETOs thus satisfy separate equations depending on the values of $l$. The centrifugal term is retained only formally. The substitution $G^{\alpha}_{nl}=r^{(1-\alpha)/2}R^{\alpha}_{nl}$ removes the first derivative in the second term and yields ordinary Coulomb$-$Sturmians with effective principal and angular momentum quantum numbers. The operator,
\begin{multline}\label{eq:s214}
H^{\alpha}
=\nabla^{2}_{2}
+\frac{1-\alpha}{r}
\left(
\frac{\partial}{\partial r}
-\frac{l}{r}
\right)
\\
+\frac{\zeta\left(2n+1-\alpha\right)}{r}
-\zeta^{2},
\end{multline}
is not symmetric with respect to the three$-$dimensional volume element. The symmetry broken by the second term is restored only by the additional factor $r^{1-\alpha}$. Combined with the weight $1/r$ of the Coulomb$-$Sturmians ($\alpha=1$), this factor yields the weight $r^{-\alpha}$ of the Hilbert space in which the $\Psi^{\alpha}-$ETOs are complete and orthonormal.

Nonrelativistic Coulomb$-$Sturmian functions form a discrete basis of analytic functions, which permits the representation of both bound and scattering nonrelativistic wave functions~\cite{4Rotenberg1962,59Avery2006,60Gasaneo2013,61Ambrosio2016}. Relativistic Coulomb$-$Sturmians should possess orthogonality and completeness, with both properties reducing smoothly to their nonrelativistic counterparts in the limit $c\to\infty$~\cite{62Grant2000,63Grant2007,64Bostock2011,65Bagci2016}. The Dirac Hamiltonian $\hat H_D$ describing an electron moving through the Coulomb potential around a fixed, point$-$like nucleus of charge $Ze$, where $e$ denotes the proton charge and $Z$ is the atomic number, possesses rotational and inversion symmetry. Rotations are generated by the total angular momentum operator $\hat{\mathbf J}=\hat{\mathbf L}+\frac12\hat{\boldsymbol\sigma}$, rather than by $\hat{\mathbf L}$ alone. The integrable structure of the Dirac$-$Coulomb problem is characterized by the complete set of commuting operators $\{\hat H_D,\hat{\mathbf J}^{2},\hat J_z,\hat K\}$. Here, $\hat K$ is the Dirac invariant~\cite{48Greiner2000,63Grant2007}. In atomic units, it is defined by $\hat K=\beta(\hat{\boldsymbol\sigma}\cdot\hat{\mathbf L}+\mathbb I)$, where $\beta$ is the Dirac matrix and $\mathbb I$ is the identity operator in spinor space. The eigenvalues of $\hat K$ take the discrete values $\kappa=\pm\left(j+\frac12\right)$, $\kappa=\pm1,\pm2,\pm3,\ldots$. For negative $\kappa$, the angular momentum quantum number is $l=-\kappa-1$, corresponding to $j=l+\frac12$. For positive values of $\kappa$ it is $l=\kappa$, corresponding to $j=l-\frac12$. The quantum number $\kappa$ identifies the spin$-$orbit coupled states of total angular momentum quantum number $j$ and fixes their parity. Consequently, any consistent rearrangement of the function $D^{p}_{q-p}$ in the radial differential equation must be compatible with a Coulomb$-$Sturmian differential equation which preserves the spin$-$orbit coupling structure of the relativistic Dirac$-$Coulomb problem in the nonrelativistic limit ($c\to\infty$).

The rearrangement of the function $D^{p}_{q-p}$ according to the aforementioned discussion should be as follows,
\begin{multline}\label{eq:s215}
D^{2l+2-\alpha}_{n-l-1}\left(x\right)
=\underline{4l^{2}+8l+4-4l\alpha-4\alpha}
+\underline{4\alpha-4}
\\
+\underline{2\alpha x-4nx-2x+x^{2}}.
\end{multline}
Eq.~\eqref{eq:s212} can thus be rewritten in the form:
\begin{multline}\label{eq:s216}
x\frac{d^{2}}{dx^{2}}
\mathcal{R}^{\alpha}_{nl}\left(x\right)
+\left(3-\alpha\right)\frac{d}{dx}
\mathcal{R}^{\alpha}_{nl}\left(x\right)
\\
+\Bigg[
n+\left(1-\alpha\right)\left(\frac{1}{x}
+\frac{1}{2}\right)
\\
-\frac{\left(l+1\right)\left(l+1-\alpha\right)}{x}
-\frac{x}{4}
\Bigg]
\mathcal{R}^{\alpha}_{nl}\left(x\right)=0.
\end{multline}
It is observed that Eq.~\eqref{eq:s216} may also be recovered from Eq.~\eqref{eq:s212} by adding and subtracting the term $(1-\alpha)/x$ within the brackets. The attempt to retain the centrifugal term discussed above, while mixing the centrifugal and Coulomb interactions, led Guseinov to treat the parameter $\alpha$ as an observable and to interpret it as a quantum number associated with a self$-$frictional force produced by the particle itself~\cite{58Guseinov2012,66Guseinov2017}.\\
Returning to the variable $r=x/\left(2\zeta\right)$ and collecting the resulting radial derivative terms, the differential equation reads:
\begin{multline}\label{eq:s217}
\Bigg[
-\frac{1}{\left(3-\alpha\right)r^{3-\alpha}}
\frac{\partial}{\partial r}
\left(
r^{3-\alpha}\frac{\partial}{\partial r}
\right)
+\frac{l\left(l+2-\alpha\right)}
{\left(3-\alpha\right)r^{2}}
\\
-\frac{\zeta\left(2n+1-\alpha\right)}
{\left(3-\alpha\right)r}-E
\Bigg]
\mathcal{R}^{\alpha}_{nl}\left(\zeta, r\right)=0.
\end{multline}
where, $\left(3-\alpha\right)E=-\zeta^{2}$. Eq.~\eqref{eq:s217}, satisfied by the $\Psi^{\alpha}-$ETOs, coincides with the $d-$dimensional Coulomb$-$Sturmian equation~\cite{67Avery2000,68Aquilanti2001,69Coletti2013,70Atawnah2025},
\begin{align}\label{eq:s218}
\Psi^{\alpha}_{nlm}\left(\zeta, \vec{r}\right)
=\mathcal{R}^{\alpha}_{nl}\left(\zeta, r\right)\mathcal{Y}_{lm}\left(\Omega_{3-\alpha}\right),
\end{align}
where $\mathcal{Y}_{lm}(\Omega_{3-\alpha})$ are hyperspherical harmonics~\cite{67Avery2000,69Coletti2013}, $m$ denoting collectively the $2-\alpha$ remaining angular quantum numbers, and,
\begin{align}\label{eq:s219}
\left[
\nabla^{2}_{3-\alpha}
+\frac{2\zeta\left(n+\frac{1-\alpha}{2}\right)}{r}
-\zeta^{2}
\right]
\Psi^{\alpha}_{nlm}\left(\zeta, \vec{r}\right)=0,
\end{align}
here,
\begin{align}\label{eq:s220}
\nabla^{2}_{3-\alpha}
=\frac{1}{r^{3-\alpha}}
\frac{\partial}{\partial r}
\left(
r^{3-\alpha}\frac{\partial}{\partial r}
\right)
-\frac{\Lambda^{2}_{3-\alpha}}{r^{2}},
\end{align}
is the $\left(4-\alpha\right)-$dimensional Laplacian with $\Lambda^{2}_{3-\alpha}$~\cite{67Avery2000,70Atawnah2025},
\begin{multline}\label{eq:s221}
\Lambda^{2}_{3-\alpha}
=-\Bigg[
\frac{1}{\sin^{2-\alpha}\theta_{1}}
\frac{\partial}{\partial\theta_{1}}
\left(
\sin^{2-\alpha}\theta_{1}
\frac{\partial}{\partial\theta_{1}}
\right)
\\
+\frac{1}{\sin^{2}\theta_{1}\sin^{1-\alpha}\theta_{2}}
\frac{\partial}{\partial\theta_{2}}
\left(
\sin^{1-\alpha}\theta_{2}
\frac{\partial}{\partial\theta_{2}}
\right)
\\
+\ldots
+\frac{1}{\sin^{2}\theta_{1}
\ldots
\sin^{2}\theta_{2-\alpha}}
\frac{\partial^{2}}{\partial\varphi^{2}}
\Bigg],
\end{multline}
is the negative of the Laplace$-$Beltrami operator on the $(3-\alpha)-$dimensional unit sphere $\mathcal S^{3-\alpha}$, with eigenvalues $l(l+2-\alpha)$. The number of spatial dimensions is $d=4-\alpha$, and $\Omega_{3-\alpha}$ denotes the angular coordinates $\theta_{1},\ldots,\theta_{2-\alpha},\varphi$ on $\mathcal{S}^{3-\alpha}$.
\subsection{BH$-$ETOs as $d-$Dimensional Coulomb$-$Sturmians with Noninteger Quantum Numbers}\label{sec:BHETOsDDim}
In the nonrelativistic limit, the formulation of the radial Dirac solution as a generalized Kepler problem by Infeld and Hull reduces to a modified form of the radial Schr{\"o}dinger equation, in which the quantum numbers are no longer restricted to integer values. The corresponding differential equation reads~\cite{47Infeld1951},
\begin{multline}\label{eq:s222}
\frac{d^{2}u_{n^{\ast}_{\epsilon}l^{\ast}_{\epsilon}}\left(r\right)}{dr^{2}}
+\frac{2}{r}u_{n^{\ast}_{\epsilon}l^{\ast}_{\epsilon}}\left(r\right)
-\frac{l^{\ast}_{\epsilon}\left(l^{\ast}_{\epsilon}+1\right)}{r^{2}}
u_{n^{\ast}_{\epsilon}l^{\ast}_{\epsilon}}\left(r\right)
\\
-\frac{1}{n^{\ast 2}_{\epsilon}}u_{n^{\ast}_{\epsilon}l^{\ast}_{\epsilon}}\left(r\right)=0,
\end{multline}
here, $u_{n^{\ast}_{\epsilon}l^{\ast}_{\epsilon}}=rR_{n^{\ast}_{\epsilon}l^{\ast}_{\epsilon}}$ are the reduced radial functions, $n^{\ast}_{\epsilon}\in\mathbb R_{\geq1}$, $n^{\ast}_{\epsilon}=n+\epsilon$, $l^{\ast}_{\epsilon}\in\mathbb R_{\ge0}$, $l^{\ast}_{\epsilon}=l+\epsilon$, $0\le\epsilon<1$. Solutions of Eq.~\eqref{eq:s222} are a variant of the Slater$-$type orbitals characterized by noninteger principal quantum numbers $n^{\ast}_{\epsilon}\ge1$, where $\frac{1}{n^{\ast}_{\epsilon}}$ is replaced by $\zeta$:
\begin{equation}\label{eq:s223}
R_{n^{\ast}_{\epsilon}l^{\ast}_{\epsilon}}\left(\zeta, r\right)
=\frac{\left(2\zeta\right)^{n^{\ast}_{\epsilon}+1/2}}
{\sqrt{\Gamma\left(2n^{\ast}_{\epsilon}+1\right)}}
r^{n^{\ast}_{\epsilon}-1}
e^{-\zeta r}.
\end{equation}
In the case of integer quantum numbers, these functions are obtained by retaining only the highest power of $r$ in the Laguerre polynomials of the hydrogen$-$like wave functions. Infeld and Hull did not explicitly report a generalization of the differential equation given in Eq.~\eqref{eq:s222}, or of its solution, to hydrogen$-$like systems. Yet, analogously to obtaining Slater$-$type orbitals from hydrogen$-$like functions, a compact form solution of Eq.~\eqref{eq:s222} which satisfies the boundary condition imposed for the radial part may be written as,
\begin{multline}\label{eq:s224}
R_{n^{\ast}_{\epsilon}l^{\ast}_{\epsilon}}\left(\zeta, r\right)
=\mathcal{N}_{n^{\ast}_{\epsilon}l^{\ast}_{\epsilon}}\left(\zeta\right)
\left(2\zeta r\right)^{l^{\ast}_{\epsilon}}
e^{-\zeta r}
\\
\times\mathcal{L}^{2l^{\ast}_{\epsilon}+1}_{n^{\ast}_{\epsilon}-l^{\ast}_{\epsilon}-1}
\left(2\zeta r\right),
\end{multline}
where $\mathcal{N}_{n^{*}_{\epsilon}l^{*}_{\epsilon}}$ is the normalization constant. Eq.~\eqref{eq:s224} is a solution of the differential equation given in Eq.~\eqref{eq:s222} for the generalized Kepler problem. It provides a proper representation of the relativistic wave functions of hydrogen$-$like systems for $n^{\ast}_{\gamma}=n_r+\gamma$, where $n_r$ is the radial quantum number~\cite{48Greiner2000} with $n_r=0,1,2,\ldots$,
\begin{align}\label{eq:s225}
\gamma=\sqrt{\kappa^{2}-\frac{Z^{2}}{c^{2}}},
\end{align}
and $\kappa$ is the eigenvalue of the Dirac spin$-$orbit operator introduced above. Each of the two terms whose sum constitutes the large and small components of the relativistic hydrogen$-$like wave function~\cite{48Greiner2000,63Grant2007} has the form of Eq.~\eqref{eq:s224}. The principal quantum number entering both terms takes noninteger values, $n^{\ast}_{\gamma}$. It arises from the condition terminating the series solution of the Dirac$-$Coulomb radial equations. In the nonrelativistic limit, $\gamma$ reduces to $|\kappa|$ and $n^{\ast}_{\gamma}$ becomes $n^{\ast}_{\gamma}=n\in\mathbb N_{>0}$, $n=n_r+|\kappa|$. The integer $n$ employed in~\cite{62Grant2000,63Grant2007} for the construction of $L-$spinors does not follow from the termination condition alone. It additionally requires replacing $\gamma$ by $|\kappa|$. From this point of view, it is therefore appropriate to first consider the Laguerre polynomials, given that they appear in the components of the relativistic wave function.

The differential equation of the generalized Laguerre polynomials with $p^{\ast}>-1$, $q^{\ast} \in \mathbb{R}$, and $(q^{*}-p^{*})\in\mathbb N_{\ge0}$ is,
\begin{multline}\label{eq:s226}
x\frac{d^{2}y\left(x\right)}{dx^{2}}
+\left(p^{\ast}+1-x\right)
\frac{dy\left(x\right)}{dx}
+\left(q^{\ast}-p^{\ast}\right)
y\left(x\right)=0.
\end{multline}
The Laguerre polynomials in Eq.~\eqref{eq:s224} have properties similar to those of the Laguerre polynomials appearing in the $L-$spinors. In both cases the degree $(q^{\ast}-p^{\ast})$ takes non$-$negative integer values. To be explicit, the values of $p^{\ast}$ and $q^{\ast}$ are $p^{\ast}=2l^{*}_{\varepsilon}+1$, $q^{*}=n^{\ast}_{\varepsilon}+l^{\ast}_{\varepsilon}$ for Eq.~\eqref{eq:s224} and $p^{\ast}=2\gamma$, $q^{\ast}=n_r+2\gamma$ (or $q^{\ast}=n_r+2\gamma-1$) with $n_r=n-|\kappa|$ ($n^{\ast}_{\gamma}=n\in\mathbb N_{>0}$) for the $L-$spinors. In the nonrelativistic limit $\{n^{\ast}_{\varepsilon},n^{\ast}_{\gamma}\}\ge1$ hold. The solutions of quantum defect theory with noninteger $n^{\ast}$ and $l^{\ast}$~\cite{45Kostelecky1985}, discussed in Section~\ref{sec:intro}, are examples of such solutions with $n^{\ast}\ge1$. The solution of the Dirac equation, however, behaves near the nucleus as $r^{\gamma}$, with $\gamma$ smaller than $\vert \kappa \vert$, and the condition that terminates the solution of the Dirac$-$Coulomb radial differential equation imposes $n^{\ast}_{\gamma}=n_r+\gamma$, which for $n_r=0$ gives $n^{\ast}_{\gamma}>0$. In the nonrelativistic limit, the solution of the Dirac$-$Coulomb equation, with $\gamma$ fixed by the nuclear charge $Z$ and by $\kappa$, restricts the principal quantum number $n^{\ast}_{\gamma}$ to values $n^{\ast}_{\gamma}=n\ge1$. Since the relativistic solution itself admits $0<n^{\ast}_{\gamma}<1$, one also expects a solution whose nonrelativistic limit applies for $n^{\ast}>0$. Quantum numbers with noninteger values were also used in numerous atomic structure calculations~\cite{37Bagci2020} based on the self$-$consistent field method. These calculations employ a variant of Slater$-$type functions different from that given in Eq.~\eqref{eq:s223}. Such functions were not derived from any differential equation. Rather, they were introduced hypothetically and are the $n^{\ast}-$STOs of Section~I. They are given as,
\begin{align}\label{eq:s227}
R_{n^{\ast}}\left(\zeta, r\right)
=\frac{\left(2\zeta\right)^{n^{\ast}+1/2}}
{\sqrt{\Gamma\left(2n^{\ast}+1\right)}}
r^{n^{\ast}-1}
e^{-\zeta r}.
\end{align}
The functions in Eq.~\eqref{eq:s223} are obtained from a hydrogen$-$like differential equation, so the subscript $l^{\ast}_{\epsilon}$ implies the restriction $n^{\ast}_{\epsilon}=l^{\ast}_{\epsilon}+1$ on the principal quantum number. The $n^{\ast}-$STOs, by contrast, are free from such a restriction and $n^{\ast}>0$. Variational calculations performed in~\cite{37Bagci2020,51Bagci2025,52Bagci2025} also demonstrate that allowing the principal quantum number to take noninteger values, occasionally falling within the range $0<n^{\ast}\le1$, improves the description of atomic orbitals. Suppose that the $n^{\ast}-$STOs are obtained from the solution of a Schr\"odinger$-$like differential equation with quantum numbers $\{n^{\ast}_{\nu},l^{\ast}_{\nu},m^{\ast}_{\nu}\}$, $0<\nu\le1$, by retaining only the term containing the highest power of $r$ in the Laguerre polynomials. Two possibilities then arise for the relation between the principal and angular momentum quantum numbers, each allowing $n^{\ast}=n^{\ast}_{\nu}>0$:
\begin{align}\label{eq:s228}
\begin{aligned}
n^{\ast}_{\nu}&\ge l+\nu,
\\
n^{\ast}_{\nu}&\ge\nu(l+1).
\end{aligned}
\end{align}
Together with the solution of the differential equation of Infeld and Hull in Eqs.~\eqref{eq:s222} and \eqref{eq:s223}, the above relations yield four variants of Slater$-$type orbitals. They were introduced by one of the authors~\cite{50Bagci2023} as:
\begin{subequations}\label{eq:s229}
\begin{align}
&R^{\alpha}_{n\,n-1} : \left[n\in\mathbb{N}_{\ge1}, l\in\left\{0,1,\ldots,n-1\right\}\right],\label{eq:s229a}
\\
&R^{\alpha}_{n^{\ast}_{\nu}n^{\ast}_{\nu}-1} : \left[n^{\ast}_{\nu}\in\mathbb{R}_{\ge1}, n^{\ast}_{\nu}\ge l+\nu+1, l^{\ast}_{\nu}=l+\nu\right],\label{eq:s229b}
\\
&R^{\alpha}_{n^{\ast}_{\nu}n^{\ast}_{\nu}-\nu} : \left[n^{\ast}_{\nu}\in\mathbb{R}_{>0}, n^{\ast}_{\nu}\ge l+\nu, l^{\ast}_{\nu}=l\right],\label{eq:s229c}
\\
&R^{\alpha}_{n^{\ast}_{\nu}n^{\ast}_{\nu}-\nu} : \left[n^{\ast}_{\nu}\in\mathbb{R}_{>0}, n^{\ast}_{\nu}\ge\nu\left(l+1\right), l^{\ast}_{\nu}=l\nu\right].\label{eq:s229d}
\end{align}
\end{subequations}
BH$-$ETOs compatible with all the variants given in Eqs.~\eqref{eq:s229a}$-$\eqref{eq:s229d} were derived in~\cite{50Bagci2023} and are given by,
\begin{multline}\label{eq:s230}
\mathcal{R}^{\alpha\nu}_{n^{\ast}_{\nu}l^{\ast}_{\nu}}\left(\zeta, r\right)
=\mathcal{N}^{\alpha\nu}_{n^{\ast}_{\nu}l^{\ast}_{\nu}}\left(\zeta\right)
\left(2\zeta r\right)^{l^{\ast}_{\nu}+\nu-1}
\\
\times 
e^{-\zeta r}
\mathcal{L}^{2l^{\ast}_{\nu}+2\nu-\alpha}_{n^{\ast}_{\nu}-l^{\ast}_{\nu}-\nu}
\left(2\zeta r\right).
\end{multline}
This compact closed form follows from the finite linear combinations that relate the BH$-$ETOs to these variants. The aforementioned wave functions are subsets of Eq.~\eqref{eq:s230}. Thus, the subscript $\nu$ is omitted henceforth, i.e., $n^{\ast}_{\nu}\equiv n^{\ast}$, $l^{\ast}_{\nu}\equiv l^{\ast}$, and $m^{\ast}_{\nu}\equiv m^{\ast}$. With $p^{\ast}=2l^{\ast}+2\nu-\alpha$ and $q^{\ast}=n^{\ast}+l^{\ast}+\nu-\alpha$, the normalization constant given in Eq.~\eqref{eq:s24} becomes,
\\
for $n^{\ast}-l^{\ast}-\nu\in\mathbb N_{0}$ and $p^{\ast}>-1$,
\begin{multline}\label{eq:s231}
\mathcal{N}^{\alpha\nu}_{n^{\ast}l^{\ast}}\left(\zeta\right)
\\
=\left[
\frac{\left(2\zeta\right)^{4-\alpha}}
{2n^{\ast}+1-\alpha}
\frac{\Gamma\left(n^{\ast}-l^{\ast}-\nu+1\right)}
{\Gamma\left(n^{\ast}+l^{\ast}+\nu+1-\alpha\right)}
\right]^{1/2}.
\end{multline}
Prior to recent developments, one of the authors, following Guseinov~\cite{54Guseinov2002}, regarded the $\Psi^{\alpha}-$ETOs as complete orthonormal sets of exponential$-$type orbitals and, accordingly, introduced the BH$-$ETOs as their generalization to noninteger quantum numbers. From this point of view, the completeness and orthonormality of these functions are defined with respect to the weighted Hilbert space $L^{2}_{r^{-\alpha}}(\mathbb R^{3})$, with each value of $\alpha$ corresponding to a distinct complete set. It has been recently shown by one of the authors~\cite{53Bagci2026} that $\alpha$ is, in fact, a dimensional parameter. This result is treated in Subsection~\ref{sec:PAETOsDDim} above and formulated explicitly in Eq.~\eqref{eq:s217}. The BH$-$ETOs thereby constitute a generalization of the three$-$dimensional Coulomb$-$Sturmians to noninteger quantum numbers. Their extension to arbitrary dimension $d$, $d=4-\alpha$, follows in direct analogy with the formulation presented.\\
Following the discussion in Subsection~\ref{sec:PAETOsDDim}, the function $D^{p^{\ast}}_{q^{\ast}-p^{\ast}}(x)$ is rearranged as,
\begin{multline}\label{eq:s232}
D^{2l^{\ast}+2\nu-\alpha}_{n^{\ast}-l^{\ast}-\nu}\left(x\right)
=\underline{4l^{\ast 2}+8l^{\ast}\nu+4\nu^{2}-4l^{\ast}\alpha-4\nu\alpha}
\\
+\underline{4\alpha-4}+\underline{2\alpha x-4n^{\ast}x-2x+x^{2}}.
\end{multline}
None of the terms in the above equation cancel, in contrast to Eq.~\eqref{eq:s215}. For $\nu\ne1$, the constant terms do not cancel each other. They combine as $4l^{\ast 2}+8l^{*}\nu+4\nu^{2}-4l^{\ast}\alpha-4\nu\alpha+4\alpha-4=4(l^{\ast}+\nu-1)(l^{\ast}+\nu+1-\alpha)=4\lambda^{\ast}(\lambda^{\ast}+2-\alpha)$, with $\lambda^{\ast}=l^{\ast}+\nu-1$, and this grouping, as established in the present study, corresponds to the $d-$dimensional Coulomb$-$Sturmians. Eq.~\eqref{eq:s215} is recovered for $\nu=1$. The dimension related terms $\alpha^{2}-\alpha^{2}$, however, are cancel in both equations. The coefficient in the above equation takes all real values through $n^{\ast}$, while the constant term, arising from the angular momentum part, satisfies $\lambda^{\ast}(\lambda^{\ast}+2-\alpha)\ge-(2-\alpha)^{2}/4$ for every real $\lambda^{\ast}$, with equality at $l^{\ast}+\nu=\alpha/2$. This is the critical bound in $d=4-\alpha$ dimensions below which the fall to the center occurs~\cite{71Frank1971}. Values below this bound cannot be reached with real $\lambda^{\ast}$. They would require complex quantum numbers. Hence, no more general form of $D^{p^{\ast}}_{q^{\ast}-p^{\ast}}$ exists for real quantum numbers, and the resulting equation represents the most general form of the Coulomb$-$Sturmian differential equation.

The radial differential equation of the BH$-$ETOs which extends the $d-$dimensional Coulomb$-$Sturmian differential equation to noninteger angular momentum and principal quantum numbers then takes the form:
\begin{multline}\label{eq:s233}
x\frac{d^{2}}{dx^{2}}
\mathcal{R}^{\alpha\nu}_{n^{\ast}l^{\ast}}\left(x\right)
+\left(3-\alpha\right)\frac{d}{dx}
\mathcal{R}^{\alpha\nu}_{n^{\ast}l^{\ast}}\left(x\right)
\\
+\Bigg[
n^{\ast}
+\left(
1
-\alpha\right)\left(\frac{1}{x}
+\frac{1}{2}
\right)
\\
-\frac{\left(l^{\ast}+\nu\right)\left(l^{\ast}+\nu-\alpha\right)}{x}
-\frac{x}{4}
\Bigg]
\mathcal{R}^{\alpha\nu}_{n^{\ast}l^{\ast}}\left(x\right)=0.
\end{multline}
Again changing the variable as $r=\frac{x}{2\zeta}$ and collecting the resulting derivatives, Eq.~\eqref{eq:s233} reads:
\begin{multline}\label{eq:s234}
\Bigg[
-\frac{1}{\left(3-\alpha\right)r^{3-\alpha}}
\frac{\partial}{\partial r}
\left(
r^{3-\alpha}\frac{\partial}{\partial r}
\right)
\\
+\frac{\left(l^{\ast}+\nu-1\right)\left(l^{\ast}+\nu+1-\alpha\right)}
{\left(3-\alpha\right)r^{2}}
\\
-\frac{\zeta\left(2n^{\ast}+1-\alpha\right)}
{\left(3-\alpha\right)r}
-E
\Bigg]
\mathcal{R}^{\alpha\nu}_{n^{\ast}l^{\ast}}\left(\zeta, r\right)=0,
\end{multline}
here, $(3-\alpha)E=-\zeta^{2}$. Replacing $l^{\ast}+\nu-1$ with $\lambda^{\ast}=l^{\ast}+\nu-1$, the expansion of the compact form of the radial differential equation of BH$-$ETOs yields,
\begin{multline}\label{eq:s235}
\frac{d^{2}}{dr^{2}}
\mathcal{R}^{\alpha\nu}_{n^{\ast}l^{\ast}}\left(\zeta, r\right)
+\frac{\left(3-\alpha\right)}{r}
\frac{d}{dr}
\mathcal{R}^{\alpha\nu}_{n^{\ast}l^{\ast}}\left(\zeta, r\right)
\\
+\Bigg[
\frac{\zeta\left(2n^{\ast}+1-\alpha\right)}{r}
-\frac{\lambda^{\ast}\left(\lambda^{\ast}+2-\alpha\right)}{r^{2}}
\\
-\zeta^{2}
\Bigg]
\mathcal{R}^{\alpha\nu}_{n^{\ast}l^{\ast}}\left(\zeta, r\right)=0.
\end{multline}
Eqs.~\eqref{eq:s234} and \eqref{eq:s235}, with the centrifugal term $\lambda^{\ast}(\lambda^{\ast}+2-\alpha)/r^{2}$, have the same form as the radial equation for the $d-$dimensional Coulomb$-$Sturmians given in Eq.~\eqref{eq:s217}. Upon separation of variables, the radial and the angular equations are coupled through the eigenvalue of the angular operator. The differential equation given above for the radial part of BH$-$ETOs fixes the eigenvalue as $\lambda^{\ast}(\lambda^{\ast}+2-\alpha)$, yet the angular operator itself remains to be determined. It is the Laplace$-$Beltrami operator $\Lambda^{2}_{3-\alpha}$ of Eq.~\eqref{eq:s221} for integer values of quantum numbers with $\lambda^{\ast}=l$. Its eigenfunctions are the hyperspherical harmonics, i.e., the restrictions to the unit sphere $\mathcal S^{3-\alpha}$ of homogeneous harmonic polynomials of degree $l$ with eigenvalue $l(l+2-\alpha)$. These eigenvalues correspond to radial solutions that are regular at the origin and behave as $r^{l}$. By contrast, the BH$-$ETOs behave as $r^{\lambda^{\ast}}$ with noninteger $\lambda^{\ast}$. Their angular part cannot be expressed in terms of the hyperspherical harmonics of Eq.~\eqref{eq:s218}.

Returning to the discussion on the Laguerre polynomials, consider the associated Laguerre polynomials with integer values of the parameters $\{q^{\ast}=q,p^{\ast}=p\}\in\mathbb N_{0}$. An ordinary Laguerre polynomial is obtained from Eq.~\eqref{eq:s226} directly by setting $p=0$. Conversely, generalized Laguerre polynomials can be expanded in ordinary ones~\cite{55Magnus1966}. For integer parameters, $\mathcal{L}^{p^{\ast}}_{q^{\ast}-p^{\ast}}\left(x\right)$ is expanded as,
\begin{align}\label{eq:s236}
L^{p}_{q-p}\left(x\right)
=\sum_{i=0}^{q-p}
\frac{\left(p\right)_{i}}{i!}
L_{q-p-i}\left(x\right).
\end{align}
This may misleadingly suggest that, for the generalized Laguerre polynomials, the ordinary ones can likewise be recovered simply by dropping $p^{\ast}\in\mathbb R_{\ge0}$. Since $q^{\ast}-p^{\ast}\in\mathbb N_{0}$, the proper way for the generalized Laguerre polynomials however, is to use Eq.~\eqref{eq:s236}. The situation is different for the fourth variant of the Slater$-$type orbitals in Eq.~\eqref{eq:s229d}, where $q^{\ast}-p^{\ast}\in\mathbb R_{\ge0}$. Eq.~\eqref{eq:s236} is no longer applicable, since its finite sum requires $q^{\ast}-p^{\ast}$ to be a nonnegative integer. In this case, $p^{\ast}=0$ takes on a special role. The resulting functions were previously referred to as \emph{transitional Laguerre polynomials}~\cite{50Bagci2023}. They are intermediate between the ordinary and the fractional Laguerre polynomials.
\section{Solution of the BH$-$ETOs differential equation}\label{sec:BHETOsDifSol}
\subsection{Power Series Representation for BH$-$ETOs}\label{sec:BHETOsSeries}
This section is devoted to the series solution of the radial differential equation derived for the BH$-$ETOs, with the aim of verifying analytically that they are exact solutions of Eq.~\eqref{eq:s234}. For the sake of completeness, the radial equation is first written with all physical constants retained explicitly, whereas the series solution is derived from the atomic unit form. Only bound states $\left(E<0\right)$ are considered. The radial differential equation then reads:
\begin{multline}\label{eq:s337}
\Bigg\{
-\frac{\hbar^{2}}{2m_{e}}
\Bigg[
\frac{1}{\left(3-\alpha\right)r^{3-\alpha}}
\frac{\partial}{\partial r}
\left(r^{3-\alpha}\frac{\partial}{\partial r}\right)
\\
-\frac{\left(l^{\ast}+\nu-1\right)\left(l^{\ast}+\nu+1-\alpha\right)}
{\left(3-\alpha\right)r^{2}}
\Bigg]
\\
-\frac{k_{e}Ze^{2}}{\left(3-\alpha\right)r}
-\tilde{E}
\Bigg\}
\mathcal{R}^{\alpha\nu}_{n^{\ast}l^{\ast}}\left(\zeta, r\right)=0,
\end{multline}
where $\tilde E=\hbar^{2}E/2m_{e}$, $k_{e}=1/4\pi\epsilon_{0}$ is the Coulomb constant and $m_{e}$ is the electron mass. Eq.~\eqref{eq:s234} is simply recovered by using the following definitions,
\begin{align}\label{eq:s338}
\zeta&=\frac{2Z}{a_{0}(2n^{\ast}+1-\alpha)},
&
E&=-\frac{\zeta^{2}}{3-\alpha},
\end{align}
where $a_{0}$ is the Bohr radius. The solution is assumed to take the form,
\begin{align}\label{eq:s339}
\mathcal{R}^{\alpha\nu}_{n^{\ast}l^{\ast}}\left(\zeta, r\right)
=\mathcal{F}^{\alpha\nu}_{n^{\ast}l^{\ast}}
\left(\zeta, r\right)e^{-\zeta r}.
\end{align}
Upon substitution of the first and second derivatives of the radial wave function, the resulting equation becomes:
\begin{multline}\label{eq:s340}
r^{2}
\mathcal{F}^{\dprime\alpha\nu}_{n^{\ast}l^{\ast}}\left(\zeta, r\right)
+r\left[\left(3-\alpha\right)
-2\zeta r\right]
\mathcal{F}^{\prime\alpha\nu}_{n^{\ast}l^{\ast}}\left(\zeta, r\right)
\\
+\big\{
\left[
\zeta
\left(2n^{\ast}+1-\alpha\right)
-\zeta\left(3-\alpha\right)
\right]r
\\
-\left(l^{\ast}+\nu-1\right)
\left(l^{\ast}+\nu+1-\alpha\right)
\big\}
\mathcal{F}^{\alpha\nu}_{n^{\ast}l^{\ast}}\left(\zeta, r\right)=0.
\end{multline}
The solution for $\mathcal F^{\alpha\nu}_{n^{\ast}l^{\ast}}$ is expressed in terms of a power series of the form,
\begin{align}\label{eq:s341}
\mathcal{F}^{\alpha\nu}_{n^{\ast}l^{\ast}}
\left(\zeta, r\right)
=r^{s}
\sum_{i=0}^{\infty}
a_{i}r^{i}.
\end{align}
Taking the derivatives of $\mathcal F^{\alpha\nu}_{n^{\ast}l^{\ast}}$, substituting them into Eq.~\eqref{eq:s340} and rearranging terms for each $i$, the following relationship is obtained,
\begin{multline}\label{eq:s342}
\left(s+i\right)
\left(s+i+2-\alpha\right)
a_{i}
\\
-\left(l^{\ast}+\nu-1\right)
\left(l^{\ast}+\nu+1-\alpha\right)
a_{i}
\\
+2\zeta\left(n^{\ast}-s-i\right)
a_{i-1}=0.
\end{multline}
The roots are determined by the lowest order term $i=0$,
\begin{align}\label{eq:s343}
\begin{aligned}
&\left(s-l^{\ast}-\nu+1\right)
\left(s+l^{\ast}+\nu+1-\alpha\right)=0,
\\
&s_{1}=l^{\ast}+\nu-1,\qquad s_{2}=
-\left(l^{\ast}+\nu+1-\alpha\right).
\end{aligned}
\end{align}
The root $s_{1}=l^{\ast}+\nu-1$ yields the solution that is the less singular one at $r=0$, which is finite for $l^{\ast}+\nu \ge 1$, and square integrable. It is therefore classified as regular. The root $s_{2}=-(l^{\ast}+\nu+1-\alpha)$ is negative for $l^{\ast}+\nu+1>\alpha$ and corresponds to a solution that diverges as $r\to0$. It is therefore classified as irregular.\\
It should be emphasized that the irregular solution cannot be ignored in general. Its divergence causes no difficulty when the origin lies outside the region of pure Coulomb potential. In scattering problems, both the regular and the irregular solutions are required to describe the asymptotic behavior of the wave function~\cite{72Taylor1972}. In quantum defect theory, the outer radial function of Eq.~\eqref{eq:s12} is a combination of the regular and irregular Coulomb functions, matched at a finite radius $r_{0}$ between the inner region and the outer region of pure Coulomb potential. The contribution of the irregular function, determined by the quantum defect, leads to the noninteger effective quantum number of Eq.~\eqref{eq:s11}~\cite{38Seaton1983,39Fu2016}. Pairs of regular and irregular solutions are likewise employed in the description of the Stark effect in Rydberg atoms~\cite{73Fano1981,74Harmin1982,75Giannakeas2016,76Ji2024}. When the effect of the core is instead represented by an additional term in the centrifugal part of the radial differential equation, the regular solution alone suffices, and the quantum numbers take noninteger values~\cite{45Kostelecky1985}. Similarly, for a nucleus of finite size, both solutions are needed to satisfy the boundary conditions at the nuclear surface~\cite{48Greiner2000,77Yerokhin2011}. Irregular solutions are further required in the analytic continuation of the Coulomb functions to complex energies and noninteger angular momenta~\cite{78Seaton2002,79Michel2007}, in the construction of Coulomb Green's functions, and in the study of bound state, resonance and virtual state poles~\cite{80Cattapan2000,81Mukhamedzhanov2010,82Parkash2023} and of Regge trajectories~\cite{83Regge1959}. A detailed investigation of the irregular solution for noninteger quantum numbers, however, is beyond the scope of the present work, and only the regular root is used. Hence, the recurrence relationship for the coefficients takes the form,
\begin{multline}\label{eq:s344}
a_{i}
=2\zeta\frac{\left(i+l^{\ast}+\nu-1-n^{\ast}\right)}
{i\left(2l^{\ast}+2\nu+i-\alpha\right)}
a_{i-1}
\\
=2\zeta\frac{\left[i-1-\left(n^{\ast}-l^{\ast}-\nu\right)\right]}
{i\left[i+\left(2l^{\ast}+2\nu-\alpha\right)\right]}
a_{i-1}.
\end{multline}
This relation is formally identical to that satisfied by the coefficients of the generalized Laguerre polynomials~\cite{55Magnus1966}. Since the coefficients $a_{i}$ multiply $r^{i}$, the factor $2\zeta$ appears in Eq.~\eqref{eq:s344}, and the series terminates at $i=n^{\ast}-l^{\ast}-\nu$. Eq.~\eqref{eq:s341} can now be written as,
\begin{align}\label{eq:s345}
\mathcal{L}^{2l^{\ast}+2\nu-\alpha}_{n^{\ast}-l^{\ast}-\nu}\left(2\zeta r\right)
=\sum_{i=0}^{n^{\ast}-l^{\ast}-\nu}a_{i}r^{i},
\end{align}
with $a_{0}=\binom{n^{\ast}+l^{\ast}+\nu-\alpha}{n^{\ast}-l^{\ast}-\nu}$, and,
\begin{multline}\label{eq:s346}
\mathcal{R}^{\alpha\nu}_{n^{\ast}l^{\ast}}\left(\zeta, r\right)
=\mathcal{N}^{\alpha\nu}_{n^{\ast}l^{\ast}}\left(\zeta\right)
\\
\times
\left(2\zeta r\right)^{l^{\ast}+\nu-1}
\mathcal{L}^{2l^{\ast}+2\nu-\alpha}_{n^{\ast}-l^{\ast}-\nu}\left(2\zeta r\right)e^{-\zeta r}.
\end{multline}
\subsection{First$-$Order Form of BH$-$ETOs Differential Equation}\label{sec:BHETOsFirstOrd}
It has recently been shown that the radial Schr{\"o}dinger equation can be transformed into a system of coupled first$-$order differential equations~\cite{84Gebremedhin2023}. The derivation is based on the Green's function of the kinetic energy operator, and the two resulting component functions are represented by a phase$-$amplitude \emph{ansatz}. The same approach can be applied analogously to obtain the BH$-$ETO wave functions numerically. The Schr{\"o}dinger$-$like form of the differential equation for the BH$-$ETOs in Eq.~\eqref{eq:s234} is expressed as,
\begin{multline}\label{eq:s347}
\nabla^{2}_{r,\lambda^{\ast}}
\mathcal{R}^{\alpha\nu}_{n^{\ast}l^{\ast}}\left(\zeta, r\right)
\\
=-\left(3-\alpha\right)
\left[E-V\left(\zeta, r\right)\right]
\mathcal{R}^{\alpha\nu}_{n^{\ast}l^{\ast}}\left(\zeta, r\right),
\end{multline}
here, $\nabla^{2}_{r,\lambda^{\ast}}$ is the radial Laplacian with the centrifugal term,
\begin{align}\label{eq:s348}
\nabla^{2}_{r,\lambda^{\ast}}
=\nabla^{2}_{r}
-\frac{\Lambda^{\ast}}{r^{2}}
=\frac{d^{2}}{dr^{2}}
+\frac{\left(3-\alpha\right)}{r}\frac{d}{dr}
-\frac{\Lambda^{\ast}}{r^{2}},
\end{align}
with $\Lambda^{\ast}=\lambda^{\ast}(\lambda^{*}+2-\alpha)$, and,
\begin{align}\label{eq:s349}
V\left(\zeta, r\right)
=-\frac{\zeta\left(2n^{\ast}+1-\alpha\right)}{\left(3-\alpha\right)r},
\end{align}
is the total potential and $E$ is the energy parameter. The Green's function of the radial operator $\nabla^{2}_{r,\lambda^{\ast}}$ is defined through the following relationship,
\begin{align}\label{eq:s350}
\nabla^{2}_{r,\lambda^{\ast}}
g\left(r, r^{\prime}\right)
=\frac{\delta\left(r-r^{\prime}\right)}
{r^{\prime 3-\alpha}},
\end{align}
\begin{align}\label{eq:s351}
g\left(r, r^{\prime}\right)
=-\frac{1}{2l^{\ast}+2\nu-\alpha}
\frac{r_{<}^{l^{\ast}+\nu-1}}
{r_{>}^{l^{\ast}+\nu+1-\alpha}}.
\end{align}
After straightforward algebra, the solution is formally determined by,
\begin{multline}\label{eq:s352}
\mathcal{R}^{\alpha\nu}_{n^{\ast}l^{\ast}}\left(\zeta, r\right)
=\frac{1}{2l^{\ast}+2\nu-\alpha}
\Bigg(
\int_{0}^{\infty}
\frac{r_{<}^{l^{\ast}+\nu-1}}
{r_{>}^{l^{\ast}+\nu+1-\alpha}}
\\
\times 
F^{\alpha}_{n^{\ast}}\left(\zeta, r^{\prime}\right)
\mathcal{R}^{\alpha\nu}_{n^{\ast}l^{\ast}}\left(\zeta, r^{\prime}\right)
r^{\prime 2-\alpha}
dr^{\prime}
\Bigg),
\end{multline}
here,
\begin{align}\label{eq:s353}
F^{\alpha}_{n^{\ast}}\left(\zeta, r^{\prime}\right)
=\left(3-\alpha\right)
r^{\prime}
\left[
E-V\left(\zeta, r^{\prime}\right)
\right].
\end{align}
Breaking the $r'$ axis into two ranges as $\left[0,r\right]\cup\left[r,\infty\right)$, Eq.~\eqref{eq:s352} is given by,
\begin{multline}\label{eq:s354}
\mathcal{R}^{\alpha\nu}_{n^{\ast}l^{\ast}}\left(\zeta, r\right)
=\frac{1}{2l^{\ast}+2\nu-\alpha}
\Bigg[
\frac{1}{r^{l^{\ast}+\nu+1-\alpha}}
\int_{0}^{r}
r^{\prime l^{\ast}+\nu-1}
\\
\times
F^{\alpha}_{n^{\ast}}\left(\zeta, r^{\prime}\right)
\mathcal{R}^{\alpha\nu}_{n^{\ast}l^{\ast}}\left(\zeta, r^{\prime}\right)r^{\prime 2-\alpha}dr^{\prime}
\\
+r^{l^{\ast}+\nu-1}
\int_{r}^{\infty}
\frac{1}{r^{\prime l^{\ast}+\nu+1-\alpha}}
\\
\times
F^{\alpha}_{n^{\ast}}\left(\zeta, r^{\prime}\right)
\mathcal{R}^{\alpha\nu}_{n^{\ast}l^{\ast}}\left(\zeta, r^{\prime}\right)
r^{\prime 2-\alpha}
dr^{\prime}
\Bigg].
\end{multline}
Eq.~\eqref{eq:s354} may also be presented in the following compact form,
\begin{multline}\label{eq:s355}
\mathcal{R}^{\alpha\nu}_{n^{\ast}l^{\ast}}\left(\zeta, r\right)
=\frac{1}{r^{l^{\ast}+\nu+1-\alpha}}
\mathcal{S}^{\alpha\nu}_{n^{\ast}l^{\ast}}\left(\zeta, r\right)
\\
+r^{l^{\ast}+\nu-1}
\mathcal{C}^{\alpha\nu}_{n^{\ast}l^{\ast}}\left(\zeta, r\right).
\end{multline}
The factor $1/\left(2l^{\ast}+2\nu-\alpha\right)$ is included in the component functions $\mathcal{S}^{\alpha\nu}_{n^{\ast}l^{\ast}}$ and $\mathcal{C}^{\alpha\nu}_{n^{\ast}l^{\ast}}$. Taking into account the first derivatives of $\mathcal{S}^{\alpha\nu}_{n^{\ast}l^{\ast}}$ and $\mathcal{C}^{\alpha\nu}_{n^{\ast}l^{\ast}}$, and after a few rearrangements, the resulting first$-$order matrix differential equation is:
\begin{multline}\label{eq:s356}
\begin{bmatrix}
\mathcal{S}^{\prime\alpha\nu}_{n^{\ast}l^{\ast}}\left(\zeta, r\right)
\\
\mathcal{C}^{\prime\alpha\nu}_{n^{\ast}l^{\ast}}\left(\zeta, r\right)
\end{bmatrix}
=\frac{F^{\alpha}_{n^{\ast}}\left(\zeta, r\right)}{2l^{\ast}+2\nu-\alpha}
\\
\times
\begin{bmatrix}
1 & r^{2l^{\ast}+2\nu-\alpha}
\\
-r^{-\left(2l^{\ast}+2\nu-\alpha\right)} & -1
\end{bmatrix}
\begin{bmatrix}
\mathcal{S}^{\alpha\nu}_{n^{\ast}l^{\ast}}\left(\zeta, r\right)
\\
\mathcal{C}^{\alpha\nu}_{n^{\ast}l^{\ast}}\left(\zeta, r\right)
\end{bmatrix},
\end{multline}
with,
\begin{align}\label{eq:s357}
\mathcal{S}^{\prime\alpha\nu}_{n^{\ast}l^{\ast}}\left(\zeta, r\right)
=-r^{2l^{\ast}+2\nu-\alpha}\mathcal{C}^{\prime\alpha\nu}_{n^{\ast}l^{\ast}}\left(\zeta, r\right).
\end{align}
Eq.~\eqref{eq:s356} can be solved by means of the phase$-$amplitude \emph{ansatz}~\cite{84Gebremedhin2023}. Numerical techniques such as adaptive step sizes are required to achieve high precision, since the phase varies rapidly near the nodes of the wave function. Note that the energy parameter $E$ has to be known in order to solve the matrix differential equation. Section~\ref{sec:BHETOsSeries} provides the solution of the differential equation of the BH$-$ETOs using the power series representation. Proceeding from that solution, closed analytical expressions for the component functions $\mathcal{S}^{\alpha\nu}_{n^{\ast}l^{\ast}}$ and $\mathcal{C}^{\alpha\nu}_{n^{\ast}l^{\ast}}$ can be found. The resulting expressions may be used to confirm that the BH$-$ETOs satisfy the first$-$order matrix differential equation and serve as an exact benchmark for the numerical solution of Eq.~\eqref{eq:s356}. Inserting the power series of~\eqref{eq:s345} into Eq.~\eqref{eq:s354}, with $(3-\alpha)E=-\zeta^{2}$, the component functions $\mathcal{S}^{\alpha\nu}_{n^{\ast}l^{\ast}}$ and $\mathcal{C}^{\alpha\nu}_{n^{\ast}l^{\ast}}$ take the form,
\begin{multline}\label{eq:s358}
\mathcal{S}^{\alpha\nu}_{n^{\ast}l^{\ast}}\left(\zeta, r\right)
=\mathcal{N}^{\alpha\nu}_{n^{\ast}l^{\ast}}\left(\zeta\right)
\frac{\left(2\zeta\right)^{l^{\ast}+\nu}}
{2\left(2l^{\ast}+2\nu-\alpha\right)}
\\
\times
r^{2l^{\ast}+2\nu-\alpha+1}
e^{-\zeta r}
\mathcal{G}^{\alpha\nu}_{n^{\ast}l^{\ast}}\left(\zeta, r\right),
\end{multline}
\begin{multline}\label{eq:s359}
\mathcal{C}^{\alpha\nu}_{n^{\ast}l^{\ast}}\left(\zeta, r\right)
=\mathcal{N}^{\alpha\nu}_{n^{\ast}l^{\ast}}\left(\zeta\right)
\left(2\zeta\right)^{l^{\ast}+\nu-1}
e^{-\zeta r}
\\
\times
\left[
\mathcal{L}^{2l^{\ast}+2\nu-\alpha}_{n^{\ast}-l^{\ast}-\nu}\left(2\zeta r\right)
-\frac{\zeta r}{2l^{\ast}+2\nu-\alpha}
\mathcal{G}^{\alpha\nu}_{n^{\ast}l^{\ast}}\left(\zeta, r\right)
\right],
\end{multline}
here,
\begin{multline}\label{eq:s360}
\mathcal{G}^{\alpha\nu}_{n^{\ast}l^{\ast}}\left(\zeta, r\right)
\\
=\mathcal{L}^{2l^{\ast}+2\nu-\alpha}_{n^{\ast}-l^{\ast}-\nu}\left(2\zeta r\right)
+2\mathcal{L}^{2l^{\ast}+2\nu-\alpha+1}_{n^{\ast}-l^{\ast}-\nu-1}\left(2\zeta r\right).
\end{multline}
The component functions in the phase$-$amplitude \emph{ansatz} are represented by,
\begin{align}
\begin{aligned}\label{eq:s361}
\mathcal{S}^{\alpha\nu}_{n^{\ast}l^{\ast}}\left(\zeta, r\right)
&
=\mathcal{A}^{\alpha\nu}_{n^{\ast}l^{\ast}}\left(\zeta, r\right)
\sin\left[
\Phi^{\alpha\nu}_{n^{\ast}l^{\ast}}\left(\zeta, r\right)
\right],
\\
\mathcal{C}^{\alpha\nu}_{n^{\ast}l^{\ast}}\left(\zeta, r\right)
&
=\mathcal{A}^{\alpha\nu}_{n^{\ast}l^{\ast}}\left(\zeta, r\right)
\cos\left[
\Phi^{\alpha\nu}_{n^{\ast}l^{\ast}}\left(\zeta, r\right)
\right].
\end{aligned}
\end{align}
The amplitude $\mathcal{A}^{\alpha\nu}_{n^{\ast}l^{\ast}}$ and the phase $\Phi^{\alpha\nu}_{n^{\ast}l^{\ast}}$ functions are calculated through the component functions as,
\begin{align}\label{eq:s362}
\begin{aligned}
\mathcal{A}^{\alpha\nu}_{n^{\ast}l^{\ast}}\left(\zeta, r\right)
&
=\sqrt{
\left[
\mathcal{S}^{\alpha\nu}_{n^{\ast}l^{\ast}}\left(\zeta, r\right)
\right]^{2}
+\left[
\mathcal{C}^{\alpha\nu}_{n^{\ast}l^{\ast}}\left(\zeta, r\right)
\right]^{2}
},
\\
\Phi^{\alpha\nu}_{n^{\ast}l^{\ast}}\left(\zeta, r\right)
&=\tan^{-1}
\left[
\frac{\mathcal{S}^{\alpha\nu}_{n^{\ast}l^{\ast}}\left(\zeta, r\right)}
{\mathcal{C}^{\alpha\nu}_{n^{\ast}l^{\ast}}\left(\zeta, r\right)}
\right].
\end{aligned}
\end{align}
\section{On the operator form of the radial differential equation}\label{sec:OptFRadDif}
In three dimensions, the radial momentum operator of a free particle is defined through quantizing the symmetrized form of the classical radial momentum~\cite{85Podolsky1928,86Dirac1958}. It was shown that the analogous procedure in cylindrical coordinates fails to reproduce the free particle physical Hamiltonian in that simple form. Instead, one obtains~\cite{87Liboff2003,88Paz2001},
\begin{align}\label{eq:s463}
\hat{H}_{\rm phys}
=\frac{\hat{P}^{2}_{\rho}}{2}
+\frac{\hat{L}^{2}_{z}}{2\rho^{2}}
+\frac{\hat{P}^{2}_{z}}{2}
-\frac{1}{8\rho^{2}},
\end{align}
with an additional term $-1/8\rho^{2}$ that does not appear in the one$-$ and three$-$dimensional cases. These discussions were also extended to the self$-$adjointness of the radial momentum operator, which was rigorously shown to be symmetric but not self$-$adjoint. It thus cannot be regarded as an observable in the strict sense~\cite{89Paz2002}. Studies for hydrogen$-$like atoms were carried out similarly in this context~\cite{90AlJaber1995,91AlJaber1997,92AlJaber1998}. The extra term for such systems becomes explicit when the radial differential equation is reduced to a one$-$dimensional form by the substitution $R(r)=r^{-(d-1)/2}u(r)$, which removes the first derivative term~\cite{88Paz2001}. Equivalently, it arises when the Hamiltonian is written in terms of the radial momentum operator defined through the symmetrized form~\cite{88Paz2001,89Paz2002,93Juric2022},
\begin{align}\label{eq:s464}
\hat{p}_{r}R\left(r\right)
=-\frac{i}{2}
\left\{
\hat{e}_{r}
\cdot\vec{\nabla}
R\left(r\right)
+\vec{\nabla}\cdot
\left[
\hat{e}_{r}R\left(r\right)
\right]
\right\}.
\end{align}
From Eq.~\eqref{eq:s464}, one finds the square of the radial momentum operator in the form,
\begin{align}\label{eq:s465}
\hat{p}^{2}_{r}
=-\left[
\nabla^{2}_{r}
+\frac{\left(3-\alpha\right)\left(1-\alpha\right)}
{4r^{2}}
\right].
\end{align}
Once Eq.~\eqref{eq:s235} is written in terms of $\lambda^{\ast}$, the substitution
\begin{align}\label{eq:s466}
\mathcal{R}^{\alpha\nu}_{n^{\ast}l^{\ast}}\left(\zeta, r\right)
=r^{-\left(3-\alpha\right)/2}
u^{\alpha\nu}_{n^{\ast}l^{\ast}}\left(\zeta, r\right),
\end{align}
eliminates the first derivative term. One obtains a one$-$dimensional Schr{\"o}dinger$-$like equation with the effective potential as,
\begin{align}\label{eq:s467}
\left[
-\frac{1}{\left(3-\alpha\right)}
\frac{d^{2}}{dr^{2}}
+V_{\rm eff}\left(\zeta, r\right)
-E
\right]
u^{\alpha\nu}_{n^{\ast}l^{\ast}}\left(\zeta, r\right)
=0,
\end{align}
where $V_{\rm eff}$ is determined by,
\begin{multline}\label{eq:s468}
V_{\rm eff}\left(\zeta, r\right)
=-\frac{\zeta\left(2n^{\ast}+1-\alpha\right)}
{\left(3-\alpha\right)r}
\\
+\frac{1}{3-\alpha}
\left[
\lambda^{\ast}
\left(\lambda^{\ast}+2-\alpha\right)
+\frac{\left(3-\alpha\right)\left(1-\alpha\right)}{4}
\right]
\frac{1}{r^{2}}.
\end{multline}
Upon replacing the second derivative of the reduced function by the square of the radial momentum operator acting on the radial function, the corresponding Hamiltonian reads:
\begin{align}\label{eq:s469}
\hat{H}
=\frac{\hat{p}^{2}_{r}}{\left(3-\alpha\right)}
+V_{\rm eff}\left(\zeta, r\right).
\end{align}
To make it explicit, the factor $1/(3-\alpha)$ in front of the bracket containing the centrifugal and dimensional terms is merely a consequence of the Coulomb$-$Sturmian normalization. It is eliminated in the physical Hamiltonian, which is presented as follows,
\begin{align}\label{eq:s470}
\hat{H}_{\rm phys}
=\frac{\hat{p}^{2}_{r}}{2}
+\frac{\left(3-\alpha\right)}{2}
V_{\rm eff}\left(\zeta, r\right).
\end{align}
Its eigenvalue is the physical energy $E_{\rm phys}=(3-\alpha)E/2=-\zeta^{2}/2$, and by Eq.~\eqref{eq:s338} the Coulomb term of Eq.~\eqref{eq:s470} is $-Z/r$.

A proper definition of the radial momentum operator and of the corresponding Hamiltonian expressed in terms of it is essential for analyzing uncertainty in multi$-$dimensional quantum systems. This uncertainty is characterized both by Heisenberg$-$like uncertainty products and by information$-$theoretic measures such as the Shannon and R{\'e}nyi entropies and the Fisher information~\cite{94Yanez1994,95Dehesa2010,96Dehesa2021,97Dehesa2024}. In this work, the expectation values of the position and radial momentum operators are evaluated in terms of the BH$-$ETOs, as the aforementioned uncertainty measures are either expressed in terms of or bounded by these values. The uncertainty measures of the BH$-$ETOs are given in Section~\ref{sec:UncertainityBHETOs}. The expectation values of the $\hat{r},\hat{r}^{2}$ and $\hat{p}_{r},\hat{p}^{2}_{r}$ operators over normalized BH$-$ETOs are respectively defined, for the position operators, as
\begin{subequations}\label{eq:s471}
\begin{align}
\langle\hat{r}\rangle
&=\int_{0}^{\infty}
\mathcal{R}^{\ast\alpha\nu}_{n^{\ast}l^{\ast}}\left(\zeta, r\right)
\hat{r}
\mathcal{R}^{\alpha\nu}_{n^{\ast}l^{\ast}}\left(\zeta, r\right)
r^{3-\alpha}dr,
\\
\langle\hat{r}^{2}\rangle
&=\int_{0}^{\infty}
\mathcal{R}^{\ast\alpha\nu}_{n^{\ast}l^{\ast}}\left(\zeta, r\right)
\hat{r}^{2}
\mathcal{R}^{\alpha\nu}_{n^{\ast}l^{\ast}}\left(\zeta, r\right)
r^{3-\alpha}
dr,
\end{align}
\end{subequations}
and for the radial momentum operators as
\begin{subequations}\label{eq:s472}
\begin{align}
\langle\hat{p}_{r}\rangle
&=\int_{0}^{\infty}
\mathcal{R}^{\ast\alpha\nu}_{n^{\ast}l^{\ast}}\left(\zeta, r\right)
\hat{p}_{r}
\mathcal{R}^{\alpha\nu}_{n^{\ast}l^{\ast}}\left(\zeta, r\right)r^{3-\alpha}dr,
\\
\langle\hat{p}^{2}_{r}\rangle
&=\int_{0}^{\infty}
\mathcal{R}^{\ast\alpha\nu}_{n^{\ast}l^{\ast}}\left(\zeta, r\right)
\hat{p}^{2}_{r}
\mathcal{R}^{\alpha\nu}_{n^{\ast}l^{\ast}}\left(\zeta, r\right)
r^{3-\alpha}
dr,
\end{align}
\end{subequations}
here, the radial momentum operator $\hat p_{r}$ is given by
\begin{align}\label{eq:s473}
\hat{p}_{r}
=-i\left(
\frac{\partial}{\partial r}
+\frac{3-\alpha}{2r}
\right)
=-i
r^{-\frac{3-\alpha}{2}}\frac{\partial}{\partial r}
\left(
r^{\frac{3-\alpha}{2}}
\right)
\end{align}
These expectation values, in contrast to previous treatments in the literature~\cite{95Dehesa2010,96Dehesa2021,97Dehesa2024,98Dong2005,99Aptekarev2010}, are readily obtained via the following transformation between the BH$-$ETOs and $d-$dimensional Slater$-$type orbitals with noninteger quantum numbers,
\begin{align}\label{eq:s474}
R^{\alpha}_{n^{\ast}}\left(\zeta, r\right)
=\frac{
\left(2\zeta\right)^{n^{\ast}+1-\frac{\alpha}{2}}
}
{
\sqrt{\Gamma\left(2n^{\ast}+2-\alpha\right)}
}
r^{n^{\ast}-1}
e^{-\zeta r},
\end{align}
and,
\begin{align}\label{eq:s475}
\mathcal{R}^{\alpha\nu}_{n^{\ast}l^{\ast}}\left(\zeta, r\right)
=\sum_{n^{\prime\ast}=l^{\ast}+\nu}^{n^{\ast}}
b^{\alpha\nu l^{\ast}}_{n^{\ast}n^{\prime\ast}}
R^{\alpha}_{n^{\prime\ast}}\left(\zeta, r\right),
\end{align}
where, $n^{\prime \ast}$ runs in unit steps. The transformation relationship is obtained as,
\begin{multline}\label{eq:s476}
b^{\alpha\nu l^{\ast}}_{n^{\ast}n^{\prime\ast}}
=\left(-1\right)^{n^{\prime\ast}-l^{\ast}-\nu}
\\
\times
\Bigg[
\frac{\Gamma\left(n^{\prime\ast}-\alpha+l^{\ast}+\nu+2\right)}
{\left(2n^{\ast}+1-\alpha\right)\Gamma\left(n^{\prime\ast}-\alpha+l^{\ast}+\nu+1\right)}
\\
\times
F_{n^{\prime\ast}-\alpha+l^{\ast}+\nu}\left(n^{\ast}-\alpha+l^{\ast}+\nu\right)
\\
\times
F_{n^{\prime\ast}-l^{\ast}-\nu}\left(n^{\ast}-l^{\ast}-\nu\right)
\\
\times
F_{n^{\prime\ast}-l^{\ast}-\nu}\left(2n^{\prime\ast}-\alpha+1\right)
\Bigg]^{1/2},
\end{multline}
where $F_{s}(n)=\binom{n}{s}$ are the binomial coefficients. With the normalization constant given in Eq.~\eqref{eq:s231}, the coefficients do not depend on $\zeta$. Following the discussion in Section~\ref{sec:BHETOsDDim}, this transformation is new and differs from those given in~\cite{50Bagci2023}, which were derived for BH$-$ETOs regarded as weighted wave functions in Hilbert space, whereas the present one corresponds to the proper interpretation of BH$-$ETOs as $d-$dimensional Coulomb$-$Sturmians. It holds for noninteger values of the quantum numbers.

The expectation values given in Eqs.~\eqref{eq:s471} and \eqref{eq:s472} are calculated in terms of overlap integrals. The overlap integrals over these orbitals are,
\begin{multline}\label{eq:s477}
S^{\alpha}_{n^{\ast}n^{\prime\ast}}\left(\zeta, \zeta^{\prime}\right)
=\int_{0}^{\infty}
R^{\alpha}_{n^{\ast}}\left(\zeta, r\right)
R^{\alpha}_{n^{\prime\ast}}\left(\zeta^{\prime}, r\right)
r^{3-\alpha}dr
\\
=\frac{\Gamma\left(n^{\ast}+n^{\prime\ast}-\alpha+2\right)}
{\sqrt{\Gamma\left(2n^{\ast}+2-\alpha\right)\Gamma\left(2n^{\prime\ast}-\alpha+2\right)}}
\\
\times
\left(\frac{2\zeta}{\zeta+\zeta^{\prime}}\right)^{n^{\ast}+1-\alpha/2}
\left(\frac{2\zeta^{\prime}}{\zeta+\zeta^{\prime}}\right)^{n^{\prime\ast}+1-\alpha/2},
\end{multline}
Since multiplication by $r^{s}$ transforms a Slater$-$type orbital of Eq.~\eqref{eq:s474} into another one with the principal quantum number shifted by $s$, the matrix elements of the position operators are expressed through the overlap integrals as,
\begin{multline}\label{eq:s478}
{}^{s}I^{\alpha}_{n^{\ast}n^{\prime\ast}}\left(\zeta, \zeta^{\prime}\right)
=\frac{1}{\left(2\zeta^{\prime}\right)^{s}}
\left[
\frac{\Gamma\left(2n^{\prime\ast}-\alpha+2s+2\right)}
{\Gamma\left(2n^{\prime\ast}-\alpha+2\right)}
\right]^{1/2}\\
\times S^{\alpha}_{n^{\ast}\,n^{\prime\ast}+s}\left(\zeta, \zeta^{\prime}\right).
\end{multline}
Similarly, using the definition of the radial momentum operator in~\eqref{eq:s473}, the integrals of its $k-$th power over the Slater$-$type orbitals are obtained in closed form as,
\begin{multline}\label{eq:s479}
{}^{k}P^{\alpha}_{n^{\ast}n^{\prime\ast}}\left(\zeta, \zeta^{\prime}\right)
=\left(-i\right)^{k}S^{\alpha}_{n^{\ast}n^{\prime\ast}}\left(\zeta, \zeta^{\prime}\right)
\\
\times
\sum_{j=0}^{k}\binom{k}{j}\left(-\zeta^{\prime}\right)^{k-j}
\frac{\Gamma\left(\left(2n^{\prime\ast}+3-\alpha\right)/2\right)}
{\Gamma\left(\left(2n^{\prime\ast}+3-\alpha\right)/2-j\right)}
\\
\times
\frac{\Gamma\left(n^{\ast}+n^{\prime\ast}+2-\alpha-j\right)}{\Gamma\left(n^{\ast}+n^{\prime\ast}+2-\alpha\right)}
\left(\zeta+\zeta^{\prime}\right)^{j}.
\end{multline}
Eq.~\eqref{eq:s479} reduces to Eq.~\eqref{eq:s477} for $k=0$.\\
The matrix elements of the position and radial momentum operators over the BH$-$ETOs reduce to finite double summations of the integrals over Slater$-$type orbitals given above. For the position operators, they read,
\begin{multline}\label{eq:s480}
\langle\hat{r}^{s}\rangle_{n^{\ast}l^{\ast},n^{\prime\ast}l^{\prime\ast}}
=\sum_{\mu^{\ast}=l^{\ast}+\nu}^{n^{\ast}}
\sum_{\mu^{\prime\ast}=l^{\prime\ast}+\nu}^{n^{\prime\ast}}
b^{\alpha\nu l^{\ast}}_{n^{\ast}\mu^{\ast}}
b^{\alpha\nu l^{\prime\ast}}_{n^{\prime\ast}\mu^{\prime\ast}}
\\
\times
{}^{s}I^{\alpha}_{\mu^{\ast}\mu^{\prime\ast}}\left(\zeta, \zeta^{\prime}\right).
\end{multline}
and, in the same way, for the $k-$th power of the radial momentum operator,
\begin{multline}\label{eq:s481}
\langle\hat{p}^{k}_{r}\rangle_{n^{\ast}l^{\ast},n^{\prime\ast}l^{\prime\ast}}
=\sum_{\mu^{\ast}=l^{\ast}+\nu}^{n^{\ast}}
\sum_{\mu^{\prime\ast}=l^{\prime\ast}+\nu}^{n^{\prime\ast}}
b^{\alpha\nu l^{\ast}}_{n^{\ast}\mu^{\ast}}
b^{\alpha\nu l^{\prime\ast}}_{n^{\prime\ast}\mu^{\prime\ast}}
\\
\times{}^{k}P^{\alpha}_{\mu^{\ast}\mu^{\prime\ast}}\left(\zeta, \zeta^{\prime}\right).
\end{multline}
\section{Uncertainty measures of BH$-$ETOs}\label{sec:UncertainityBHETOs}
The expectation values of Eqs.~\eqref{eq:s480} and~\eqref{eq:s481} determine the uncertainty measures of the BH$-$ETOs. The uncertainty products and the Fisher information follow in closed form, while the Shannon entropy contains one term that is evaluated numerically. Following~\cite{95Dehesa2010,96Dehesa2021,97Dehesa2024}, the radial uncertainty product, the Heisenberg uncertainty product, the Shannon entropy and the Fisher information are considered.\\
For a single Slater$-$type orbital, imposing $n^{\prime\ast}=n^{\ast}$, $\zeta^{\prime}=\zeta$ and denoting $\eta=2n^{\ast}-\alpha$,, the uncertainties of the radial position and of the radial momentum reduce to,
\begin{align}\label{eq:s482}
\begin{aligned}
\Delta r
&=\sqrt{\langle\hat{r}^{2}\rangle-\langle\hat{r}\rangle^{2}}
=\frac{\sqrt{\eta+2}}{2\zeta},
\\
\Delta p_{r}
&=\sqrt{\langle\hat{p}^{2}_{r}\rangle-\langle\hat{p}_{r}\rangle^{2}}
=\frac{\zeta}{\sqrt{\eta}},
\end{aligned}
\end{align}
\begin{align}\label{eq:s483}
\Delta r\Delta p_{r}
=\frac{1}{2}
\sqrt{\frac{\eta+2}{\eta}}
>\frac{1}{2},
\end{align}
which holds for $\eta>0$, is independent of $\zeta$ and satisfies the bound $\frac12$ that follows from $[\hat{r},\hat{p}_{r}]=i$.\\
Because the expectation values over BH$-$ETOs follow from double summations, it is more convenient to represent the uncertainty product of the radial position and radial momentum operators in matrix form. Let $\mathbf{b}$ denote the column vector with components $b^{\alpha\nu l^{\ast}}_{n^{\ast}\mu^{\ast}}$, and let $\mathbf{S}$, ${}^{s}\mathbf{I}$ and ${}^{k}\mathbf{P}$ be the matrices with elements $S^{\alpha}_{\mu^{\ast}\mu^{\prime\ast}}\left(\zeta, \zeta\right)$, ${}^{s}I^{\alpha}_{\mu^{\ast}\mu^{\prime\ast}}\left(\zeta, \zeta\right)$ and ${}^{k}P^{\alpha}_{\mu^{\ast}\mu^{\prime\ast}}\left(\zeta, \zeta\right)$, respectively. The normalization of the BH$-$ETOs then reads $\mathbf{b}^{\intercal}\mathbf{S}\mathbf{b}=1$, and the expectation values are,
\begin{align}\label{eq:s484}
\langle\hat{r}^{s}\rangle
&=\mathbf{b}^{\intercal}\:
{}^{s}\mathbf{I}\:
\mathbf{b},
&
\langle\hat{p}^{k}_{r}\rangle
&=\mathbf{b}^{\intercal}\:
{}^{k}\mathbf{P}\:
\mathbf{b},
&
k=1,2.
\end{align}
Since $\langle\hat{p}_{r}\rangle=0$, the radial uncertainty product is,
\begin{multline}\label{eq:s485}
\Delta r\Delta p_{r}
=\Big\{
\left[
\mathbf{b}^{\intercal}\:
{}^{2}\mathbf{I}\:
\mathbf{b}
-\left(
\mathbf{b}^{\intercal}\:
{}^{1}\mathbf{I}\:
\mathbf{b}
\right)^{2}
\right]
\\
\times
\mathbf{b}^{\intercal}\:
{}^{2}\mathbf{P}\:
\mathbf{b}
\Big\}^{1/2}
\ge
\frac{1}{2}.
\end{multline}

With the eigenvalue $\lambda^{\ast}\left(\lambda^{\ast}+2-\alpha\right)$ of the angular operator, Eqs.~\eqref{eq:s220} and \eqref{eq:s465} give the square of the momentum operator,
\begin{align}\label{eq:s486}
\hat{p}^{2}
=\hat{p}^{2}_{r}+\left[\lambda^{\ast}\left(\lambda^{\ast}+2-\alpha\right)+\frac{\left(3-\alpha\right)\left(1-\alpha\right)}{4}\right]\frac{1}{r^{2}},
\end{align}
where the bracketed coefficient may also be written as $\left[\left(2l^{\ast}+2\nu-\alpha\right)^{2}-1\right]/4$. In the matrix representation introduced above, the expectation value of $\hat{p}^{2}$ follows as,
\begin{multline}\label{eq:s487}
\langle\hat{p}^{2}\rangle
=\mathbf{b}^{\intercal}\:
{}^{2}\mathbf{P}\:
\mathbf{b}
\\
+\frac{\left(2l^{\ast}+2\nu-\alpha\right)^{2}-1}{4}
\mathbf{b}^{\intercal}\:
{}^{-2}\mathbf{I}\:
\mathbf{b},
\end{multline}
and the Heisenberg uncertainty product is,
\begin{multline}\label{eq:s488}
\langle\hat{r}^{2}\rangle\langle\hat{p}^{2}\rangle
\\
=\mathbf{b}^{\intercal}\:
{}^{2}\mathbf{I}\:
\mathbf{b}
\Bigg[
\mathbf{b}^{\intercal}\:
{}^{2}\mathbf{P}\:
\mathbf{b}
+\frac{\left(2l^{\ast}+2\nu-\alpha\right)^{2}-1}{4}
\mathbf{b}^{\intercal}\:
{}^{-2}\mathbf{I}\:
\mathbf{b}
\Bigg]
\\
\ge\left(\lambda^{\ast}+2-\frac{\alpha}{2}\right)^{2},
\end{multline}
which, for $\lambda^{\ast}=l$, is the bound $\left(l+2-\alpha/2\right)^{2}$ of $d-$dimensional central potentials~\cite{96Dehesa2021}.
\subsection{Shannon entropy}\label{sec:SEntropy}
The Shannon entropy~\cite{100Shannon1948} is a global measure of the spreading of the position density. It does not refer to a particular point of the domain but estimates the total extent of the density. A larger value corresponds to a more delocalized density~\cite{96Dehesa2021,97Dehesa2024,100Shannon1948}. For a central potential, the position density factorizes into radial and angular parts,
\begin{align}\label{eq:s489}
\rho\left(\zeta, \vec{r}\right)
=\vert
\mathcal{R}^{\alpha\nu}_{n^{\ast}l^{\ast}}\left(\zeta, r\right)
\vert^{2}
\vert
\mathcal{Y}_{lm}\left(\Omega_{3-\alpha}\right)
\vert^{2},
\end{align}
and the Shannon entropy decomposes accordingly,
\begin{multline}\label{eq:s490}
S\left[\rho\right]
=-\int_{0}^{\infty}\int_{\Omega_{3-\alpha}}
\rho\left(\zeta, \vec{r}\right)
\ln\rho\left(\zeta, \vec{r}\right)
r^{3-\alpha}drd\Omega_{3-\alpha}
\\
=S\left[\mathcal{R}^{\alpha\nu}_{n^{\ast}l^{\ast}}\right]
+S\left[\mathcal{Y}_{lm}\right],
\end{multline}
here,
\begin{multline}\label{eq:s491}
S\left[
\mathcal{R}^{\alpha\nu}_{n^{\ast}l^{\ast}}
\right]
\\
=-\int_{0}^{\infty}
\vert
\mathcal{R}^{\alpha\nu}_{n^{\ast}l^{\ast}}\left(\zeta, r\right)
\vert^{2}
\ln\vert
\mathcal{R}^{\alpha\nu}_{n^{\ast}l^{\ast}}\left(\zeta, r\right)
\vert^{2}
r^{3-\alpha}dr,
\end{multline}
\begin{multline}\label{eq:s492}
S\left[
\mathcal{Y}_{lm}
\right]
\\
=-\int_{\Omega_{3-\alpha}}
\vert
\mathcal{Y}_{lm}\left(\Omega_{3-\alpha}\right)
\vert^{2}
\ln\vert
\mathcal{Y}_{lm}\left(\Omega_{3-\alpha}\right)
\vert^{2}
d\Omega_{3-\alpha}.
\end{multline}
The angular part $\mathcal{Y}\left(\Omega_{3-\alpha}\right)$ is specified by the hyperspherical harmonics of Eq.~\eqref{eq:s218} only for integer $l$. For noninteger $\lambda^{\ast}$, no corresponding angular eigenfunctions exist, see Section~\ref{sec:BHETOsDDim}. The results for the BH$-$ETOs are therefore the radial parts $S\left[\mathcal{R}^{\alpha\nu}_{n^{\ast}l^{\ast}}\right]$ and $F\left[\mathcal{R}^{\alpha\nu}_{n^{\ast}l^{\ast}}\right]$, while $S\left[\mathcal{Y}\right]$ and $F\left[\mathcal{Y}\right]$ refer to integer $l$. Substituting Eq.~\eqref{eq:s230} into the expression for $S\left[\mathcal{R}^{\alpha\nu}_{n^{\ast}l^{\ast}}\right]$, the radial part decomposes into four terms,
\begin{multline}\label{eq:s493}
S\left[\mathcal{R}^{\alpha\nu}_{n^{\ast}l^{\ast}}\right]\\
=-\ln\left[\mathcal{N}^{\alpha\nu}_{n^{\ast}l^{\ast}}\left(\zeta\right)\right]^{2}
\int_{0}^{\infty}\left[\mathcal{R}^{\alpha\nu}_{n^{\ast}l^{\ast}}\left(\zeta, r\right)\right]^{2}r^{3-\alpha}dr\\
-2\lambda^{\ast}\int_{0}^{\infty}\left[\mathcal{R}^{\alpha\nu}_{n^{\ast}l^{\ast}}\left(\zeta, r\right)\right]^{2}
\ln\left(2\zeta r\right)r^{3-\alpha}dr\\
+2\zeta\int_{0}^{\infty}\left[\mathcal{R}^{\alpha\nu}_{n^{\ast}l^{\ast}}\left(\zeta, r\right)\right]^{2}r\,r^{3-\alpha}dr\\
-\int_{0}^{\infty}\left[\mathcal{R}^{\alpha\nu}_{n^{\ast}l^{\ast}}\left(\zeta, r\right)\right]^{2}\\
\times\ln\left[\mathcal{L}^{2l^{\ast}+2\nu-\alpha}_{n^{\ast}-l^{\ast}-\nu}\left(2\zeta r\right)\right]^{2}r^{3-\alpha}dr.
\end{multline}
Since the BH$-$ETOs are normalized, the first integral is unity and the remaining integrals are expectation values,
\begin{multline}\label{eq:s494}
S\left[
\mathcal{R}^{\alpha\nu}_{n^{\ast}l^{\ast}}
\right]
=-\ln\left[
\mathcal{N}^{\alpha\nu}_{n^{\ast}l^{\ast}}\left(\zeta\right)
\right]^{2}
+2\zeta\langle\hat{r}\rangle
-2\lambda^{\ast}\langle\ln\left(2\zeta r\right)\rangle
\\
-\left\langle\ln\left[
\mathcal{L}^{2l^{\ast}+2\nu-\alpha}_{n^{\ast}-l^{\ast}-\nu}\left(2\zeta r\right)
\right]^{2}\right\rangle,
\end{multline}
where $\left\langle r\right\rangle$ is given by Eq.~\eqref{eq:s480}. The expectation value $\left\langle\ln\left(2\zeta r\right)\right\rangle$ is also obtained through the transformation relationship of Eq.~\eqref{eq:s475},
\begin{multline}\label{eq:s495}
\left\langle\ln\left(2\zeta r\right)\right\rangle
=\sum_{\mu^{\ast}=l^{\ast}+\nu}^{n^{\ast}}
\sum_{\mu^{\prime\ast}=l^{\ast}+\nu}^{n^{\ast}}
b^{\alpha\nu l^{\ast}}_{n^{\ast}\mu^{\ast}}
b^{\alpha\nu l^{\ast}}_{n^{\ast}\mu^{\prime\ast}}
\\
\times S^{\alpha}_{\mu^{\ast}\mu^{\prime\ast}}\left(\zeta,\zeta\right)
\psi\left(\mu^{\ast}+\mu^{\prime\ast}+2-\alpha\right),
\end{multline}
with $\psi\left(x\right)$ is the digamma function~\cite{55Magnus1966}. The last term of Eq.~\eqref{eq:s494}, the logarithmic expectation value involving the generalized Laguerre polynomials, has no closed form in general and must be evaluated numerically.
\subsection{Fisher information}\label{sec:FEntropy}
The Fisher information~\cite{101Fisher1925} of the position density is defined by~\cite{96Dehesa2021,97Dehesa2024,101Fisher1925},
\begin{align}\label{eq:s496}
F\left[\rho\right]
=\int_{0}^{\infty}\int_{\Omega_{3-\alpha}}
\frac{
\vert\vec{\nabla}\rho\left(\zeta,\vec{r}\right)
\vert^{2}
}
{\rho\left(\zeta,\vec{r}\right)}
r^{3-\alpha}dr
d\Omega_{3-\alpha}.
\end{align}
It decomposes into radial and angular parts using,
\begin{multline}\label{eq:s497}
\frac{\vert
\vec{\nabla}\rho\left(\zeta,\vec{r}\right)
\vert^{2}
}
{\rho\left(\zeta,\vec{r}\right)}
=\frac{4\rho\left(\zeta,\vec{r}\right)
\vert
\vec{\nabla}\sqrt{\rho\left(\zeta,\vec{r}\right)}
\vert^{2}}
{\rho\left(\zeta,\vec{r}\right)}
\\
=4
\vert
\vec{\nabla}\sqrt{\rho\left(\zeta,\vec{r}\right)}
\vert^{2},
\end{multline}
\begin{multline}\label{eq:s498}
\vert
\vec{\nabla}\sqrt{\rho\left(\zeta,\vec{r}\right)}
\vert^{2}
=\left(
\frac{d\mathcal{R}^{\alpha\nu}_{n^{\ast}l^{\ast}}}{dr}
\right)^{2}
\vert
\mathcal{Y}_{lm}
\vert^{2}
\\
+\frac{
\left[
\mathcal{R}^{\alpha\nu}_{n^{\ast}l^{\ast}}
\right]^{2}
}{r^{2}}
\vert
\nabla_{\Omega}
\vert
\mathcal{Y}_{lm}
\vert
\vert^{2},
\end{multline}
and
\begin{align}\label{eq:s499}
F\left[\rho\right]
=F\left[
\mathcal{R}^{\alpha\nu}_{n^{\ast}l^{\ast}}
\right]
+\left\langle r^{-2}\right\rangle
F\left[
\mathcal{Y}_{lm}
\right],
\end{align}
with,
\begin{align}\label{eq:s4100}
F\left[
\mathcal{R}^{\alpha\nu}_{n^{\ast}l^{\ast}}
\right]
=4\int_{0}^{\infty}
\left[
\frac{d}{dr}
\mathcal{R}^{\alpha\nu}_{n^{\ast}l^{\ast}}\left(\zeta,r\right)
\right]^{2}
r^{3-\alpha}dr,
\end{align}
\begin{align}\label{eq:s4101}
F\left[
\mathcal{Y}_{lm}
\right]
=4\int_{\Omega_{3-\alpha}}
\vert
\nabla_{\Omega}\left|
\mathcal{Y}_{lm}\left(\Omega_{3-\alpha}\right)
\right|
\vert^{2}
d\Omega_{3-\alpha}.
\end{align}
Integrating the radial part by parts and using Eqs.~\eqref{eq:s465}, \eqref{eq:s471} and \eqref{eq:s472}, it is expressed as,
\begin{multline}\label{eq:s4102}
F\left[
\mathcal{R}^{\alpha\nu}_{n^{\ast}l^{\ast}}
\right]
=4\left\langle
\hat{p}^{2}_{r}
\right\rangle
+\left(3-\alpha\right)
\left(1-\alpha\right)
\left\langle r^{-2}\right\rangle
\\
=4\zeta^{2}
\left[
1-
\frac{4\lambda^{\ast}\left(\lambda^{\ast}+2-\alpha\right)}
{\left(2l^{\ast}+2\nu-\alpha\right)\left(2n^{\ast}+1-\alpha\right)}
\right].
\end{multline}
\section{Conclusions}\label{sec:Conclusions}
Unlike Eq.~\eqref{eq:s215}, none of the terms of $D^{p^{\ast}}_{q^{\ast}-p^{\ast}}$ in Eq.~\eqref{eq:s232} cancel, and its constant terms combine into $\lambda^{\ast}\left(\lambda^{\ast}+2-\alpha\right)$. The noninteger quantum numbers thus enter the radial problem only through $\lambda^{\ast}=l^{\ast}+\nu-1$, while $\alpha$ remains a dimensional parameter. Since the lower bound of $\lambda^{\ast}\left(\lambda^{\ast}+2-\alpha\right)$ coincides with the onset of the fall to the center, no more general form of $D^{p^{\ast}}_{q^{\ast}-p^{\ast}}$ exists for real quantum numbers. The resulting equation is therefore the most general form of the Coulomb$-$Sturmian differential equation for real quantum numbers.\\
The angular part shows a consequence, which is not apparent form the closed from alone. $\lambda^{\ast}$ is noninteger, the eigenvalue $\lambda^{\ast}\left(\lambda^{\ast}+2-\alpha\right)$ does not belong to the spectrum of the Laplace$-$Beltrami operator on the unit sphere. The BH$-$ETOs are thus complete radial functions without a corresponding set of angular eigenfunctions. The construction of such angular functions, and of the complete separable solution in $d$ dimensions, remains open. The same holds for the irregular solution with the root $s_{2}$. Although it is needed in scattering and in quantum defect theory, it is excluded in the present study.

The derived transformation coefficients are independent of $\zeta$. Every expectation value of the position and radial momentum operators reduces to a finite sum of gamma functions. Such a sum is meaningful only if the corresponding integral exists, which is not guaranteed for the radial momentum operator.\\
BH$-$ETOs behave as $r^{\lambda^{\ast}}$, with $\lambda^{\ast}=l^{\ast}+\nu-1$. Each application of $\hat{p}_{r}$ lowers this power by one, so that $\hat{p}^{k}_{r}\mathcal{R}^{\alpha\nu}_{n^{\ast}l^{\ast}}$ behaves as $r^{\lambda^{\ast}-k}$. The integrand of $\left\langle\hat{p}^{k}_{r}\right\rangle$, which includes the volume element $r^{3-\alpha}$, therefore behaves as $r^{2\lambda^{\ast}+3-\alpha-k}$ near the origin. The integral converges only if this power is larger than $-1$, that is, only if $k<2l^{\ast}+2\nu+2-\alpha$. For BH$-$ETOs with noninteger quantum numbers, this condition is both necessary and sufficient. The expectation value of the $k$th power of the radial momentum operator thus exists only for a finite number of powers, and this number decreases as $l^{\ast}+\nu$ decreases.\\
For integer quantum numbers, the coefficient of the divergent term contains the factor $\Gamma\left(a+1\right)/\Gamma\left(a+1-j\right)$ of Eq.~\eqref{eq:s479}, with $a=\left(2l^{\ast}+2\nu+1-\alpha\right)/2$. This factor vanishes when $a$ is a nonnegative integer, so that higher powers may remain finite, as for the hydrogenic $s$ states in $d=3$. For noninteger quantum numbers, this factor does not vanish, and the divergence is genuine. The expectation values used in the present work, those with $k=1$ and $k=2$, require $2l^{\ast}+2\nu-\alpha>0$. This is the condition under which the radial uncertainty product and the Fisher information of the BH$-$ETOs are finite. Notice that, the radial momentum operator is not self$-$adjoint~\cite{89Paz2002,93Juric2022}, its boundary terms vanish for the BH$-$ETOs with $2l^{\ast}+2\nu-\alpha>0$. This justifies $\left\langle\hat{p}_{r}\right\rangle=0$ and $\Delta r\Delta p_{r}\geq1/2$. The Heisenberg relation involves the self-adjoint operator $\hat{p}^{2}$ and is not affected.

Future work concerns the symmetry properties of the BH$-$ETOs. Their radial part retains the $\mathfrak{so}\left(2,1\right)$ algebraic structure of the Coulomb$-$Sturmians for noninteger quantum numbers, and the eigenvalue of the Casimir operator of this algebra coincides with the centrifugal coefficient of Eq.~\eqref{eq:s486}. The $\mathfrak{so}\left(d+1\right)$ symmetry of the complete problem requires hyperspherical harmonics, which do not exist for noninteger quantum numbers. The construction of the corresponding generators and Casimir operators, and the determination of the symmetry of the complete problem, remain open.
\begin{acknowledgments}
This work was supported by the Pamukkale University Scientific Research Projects Coordination Unit (PA{\"U} BAP) under Project No. 2026FEBE028.
\end{acknowledgments}

\end{document}